\documentclass[%
 reprint,
 nofootinbib,
 superscriptaddress,
 amsmath,amssymb,
 aps,
]{revtex4-2}
\usepackage{graphicx}
\usepackage{dcolumn}
\usepackage{bm}
\usepackage{lipsum}
\usepackage{cleveref}
\usepackage{amsmath,color}
\usepackage{tabularx}
\usepackage{comment}
\usepackage{caption}
\usepackage{subcaption}
\usepackage[dvipsnames]{xcolor} 

\begin{document}

\title{Training neural networks with structured noise improves classification and generalization}

\author{Marco Benedetti}
 \thanks{The two authors contributed equally}
\affiliation{
 Dipartimento di Fisica, Sapienza Università di Roma, P.le A. Moro 2, 00185 Roma, Italy
}

\author{Enrico Ventura}
\thanks{Corresponding author: enrico.ventura@uniroma1.it}
\affiliation{
 Dipartimento di Fisica, Sapienza Università di Roma, P.le A. Moro 2, 00185 Roma, Italy
 }
\affiliation{Laboratoire de Physique de l'Ecole Normale Sup\'erieure, ENS, Universit\'e PSL, F-75005 Paris, France
}

\begin{abstract}
The beneficial role of noise-injection in learning is a consolidated concept in the field of artificial neural networks, suggesting that even biological systems might take advantage of similar mechanisms to optimize their performance. 
The training-with-noise algorithm proposed by Gardner and collaborators is an emblematic example of a noise-injection procedure in recurrent networks, which can be used to model biological neural systems. 
We show how adding structure to noisy training data can substantially improve the algorithm performance, allowing the network to approach perfect retrieval of the memories and wide basins of attraction, even in the scenario of maximal injected noise. 
We also prove that the so-called Hebbian Unlearning rule coincides with the training-with-noise algorithm when noise is maximal and data are stable fixed points of the network dynamics.  
\end{abstract}

\maketitle

\section{\label{sec:intro}Introduction}
The development of models and tools from theoretical physics has greatly contributed to our understanding of how the brain is able to learn concepts \cite{amit_modeling_nodate}.
An initial approach to brain modeling used recurrent networks composed of binary neurons, connected through pairwise interactions. Each neuron could be either \textit{active} or \textit{silent}, based on the signals received from its neighbors. By incorporating realistic rules for updating neuron states (i.e. the neural dynamics), these networks exhibited dynamic multistability, resulting in the formation of many attractors. Such systems, called \textit{attractor neural networks}, are of considerable interest for multiple reasons: they can be modeled through the tools of statistical mechanics \cite{mezard_spin_1986}; they constitute a simplified version of modern artificial neural networks, and yet they present a resemblance with biological neuronal systems.\\
The landscape of attractors is shaped by both the dynamics and choice of couplings between neurons. 
The process of modification of the interactions, necessary to guarantee that concepts presented to the network are stored as attractors, is studied in neuroscience under the name of \textit{synaptic plasticity}. It is often modeled with discrete time \textit{learning algorithms}, that see a set of \textit{training data} (i.e. the concepts themselves or some representations of them) and update the interactions to reinforce memory retrieval. Modern artificial neural networks, based on deep and non-recurrent architectures, exploit the same principle to train connections between neurons from a data-set and perform a task (e.g. classification, generation of new data). \\

Attractor neural networks, as associative memory devices, share numerous formal analogies with classifiers from statistical learning.  
An important notion in this context is \textit{overfitting}, namely when a model becomes too focused on the details of the specific training set, instead of understanding the broader structure highlighted by the data. In attractor neural networks, an overfitting model would translate into a system which is not able to associate corrupted versions of a concept to the concept itself. In this case, one says that \textit{generalization} to unseen samples is lacking in the system. 
Generalization can be improved by means of regularization techniques, such as  injecting noise in the training process. Data-augmentation procedures \cite{ shorten_survey_2019, zhao_maximum-entropy_2020}, in particular, act on the training data themselves, by performing transformations (e.g. translating, rotating, blurring) to increase the heterogeneity of the data-set. Even if such procedures are practical and successful in the matter of deep networks, there are only few examples of theory-based criteria for choosing how to engineer noise to inject in the training process \cite{achille_information_2018, zhao_maximum-entropy_2020}. 
Our work considers a simple learning algorithm for attractor neural networks that employs noise-injection during training. In the context of attractor neural networks, \textit{noise-injection} refers to a random alteration of the neuronal activity encoding the concepts to be learned. 
The objective of our study is twofold: establishing a theoretical criterion for generating the most effective noise to be utilized in training, thereby optimizing the performance of an associative memory model; 
showing that injecting a maximal amount of noise, subject to specific constraints, turns the learning process into an unsupervised procedure, which is faster and more biologically plausible. Specifically, we prove that when noise is maximal, our algorithm converges to Hebbian Unlearning, a well known unsupervised learning rule \cite{hopfield_unlearning_1983}. \\

Section~\ref{sec:the_models} presents the neural network model considered in this work, as well as some relevant learning algorithms for training this architecture. \\
Section~\ref{sec:twn} introduces and characterizes the \textit{training-with-noise} algorithm \cite{gardner_training_1989}, emphasizing its relation with Support Vector Machines and with the optimization problem studied by Wong and Sherrington \cite{wong_neural_1993}. \\
Section~\ref{sec:choice_of_training_conf} contains our main results. We derive an analytical characterization of optimal noisy training data, focusing on the extreme case of injecting maximal noise. We demonstrate that generating optimal noise translates into producing training data-points whose features adhere to specific constraints, which we refer to as the \textit{structure} of the noise. When the level of noise is maximal, minima and low saddle points in the energy landscape defined by Hebb's learning rule are shown to be good training configurations according to our analytical characterization. \\
In Section~\ref{sec:HU_TWN} we show that, when noise is maximal (and our theoretical criterion applies), the Hebbian Unlearning rule emerges from the training-with-noise procedure. In this setting, the famously good performance of Hebbian Unlearning acts as a confirmation of our analytical characterization of effective training configurations. 

\section{The model}
\label{sec:the_models}
We consider a fully connected network of $N$ binary neurons $\{S_i = \pm 1\}$, $i\,\in [1,..,N]$, linked by couplings $J_{ij}$.  The network is endowed with a dynamics 
\begin{equation}
    \label{eq:dynamics}
    S_i(t+1) = \text{sign}\left(\sum_{j=1}^N J_{ij}S_j(t)\right), \hspace{0.7cm} i = 1,..,N
\end{equation}
which can be run either in parallel (i.e. \textit{synchronously}) or in series (i.e. \textit{asynchronously} in a random order) over the $i$ indices. We will mainly concentrate on asynchronous dynamics, in which case \cref{eq:dynamics} can only converge to fixed points, when they exist \cite{amit_modeling_nodate}. This kind of network can be used as an associative memory device, i.e. a system that is capable of reconstructing a number $p$ of \textit{memories} $\{\xi^{\mu}_i\}=\pm 1$, $\mu\,\in [1,...,p]$ when the dynamics is initialized to configurations that resemble them. 
Such neural networks can be studied through the tools of statistical mechanics \cite{amit_storing_1985, mezard_spin_1986, gardner_structure_1986, wong_optimally_1990}, and they can give relevant insights from the biological point of view \cite{hebb_organization_1949, amit_modeling_nodate}.\\
In this work, we will concentrate on i.i.d. memories, generated with a probability $P(\xi^{\mu}_i=\pm1)= 1/2$. Notice that memories are binary, as the network states are. 
With an appropriate choice of the couplings, the model can store an extensive number of memories $p = \alpha N$, where $\alpha$ is called \textit{load} of the network. The network performance can be benchmarked in terms of the dynamic stability achieved by the memory vectors, and the ability of the system to retrieve blurry examples of the latter. We define $\textit{classification}$ as the capability to perfectly retrieve each memory when the dynamics is initialized to the memory itself. It is convenient to define, for all site index $i$ and memory index $\mu$, the \textit{stability}
\begin{equation}
    \label{eq:stab}
    \small
    \Delta_i^{\mu} = \frac{\xi_i^{\mu}}{\sqrt{N}\sigma_i}\sum_{j = 1}J_{ij}\xi_j^{\mu}, \qquad \sigma_i = \sqrt{\sum_{j=1}^N J_{ij}^2/N}.
\end{equation}
When for a given $\mu$ all $\Delta_i^{\mu},\,i\in[1,...,N]$ are positive, $\vec{\xi}^\mu$ is \textit{correctly classified}, i.e. it is a fixed point of \cref{eq:dynamics}.
We define $\textit{generalization}$ as the capability to retrieve the memory, or a configuration that is strongly related to it, by initializing the dynamics on a corrupted version of the memory. This property of the neural network is related to the size of the basins of attraction to which the memories belong, and does not imply classification. A good measure of the performance in this sense is the \textit{retrieval map}
\begin{equation}
\label{eq:m}
m_f(m_0) := \overline{\Big\langle\frac{1}{N}\sum_{i = 1}^N \xi_i^\mu S_i^{\mu}(\infty)\Big\rangle},
\end{equation}
$i.e.$ the overlap between memory $\vec{\xi}^{\mu}$ and $\vec{S}^{\mu}(\infty)$, where $\vec{S}^{\mu}(\infty)$ is a fixed point, reached initializing the dynamics in a state with overlap $m_0$ with $\vec{\xi}^\mu$. For the rest of this work, the symbol $\overline{\hspace{0.1cm}\cdot\hspace{0.1cm}}$ will denote the average over different realizations of the memories and $\langle \cdot \rangle$ the average over different realizations of the initial state. In the classification regime, one obtains $m_f = 1$ when $m_0 = 1$.\\ 

We now provide for some notable examples of learning rules implemented on recurrent neural networks, which will often be mentioned in the rest of this work. All these procedures can be conceived as an iterative modification of the couplings on the basis of an initial choice of $J$ and a set of training data presented to the network. 

\subsection{Linear Perceptron \& Support Vector Machine (SVM) \label{sec:svm}}
Given the fully connected architecture, we want to find a set of couplings that satisfies the constraints 
\begin{equation}
\label{eq:const}
    \Delta_i^{\mu} > k , \qquad \forall \mu, i \ ,
\end{equation}
with $k$ being a positive parameter. 
This problem can be solved by the linear perceptron algorithm developed by \cite{gardner_space_1988, gardner_phase_1989, marvin_minsky_seymour_papert_perceptrons_1969}, an iterative procedure leading to dynamically stable memories and tunable generalization performance. Specifically, we are going to refer to \textit{linear perceptron} as the adaptation of the classical perceptron to recurrent neural networks, introduced in \cite{battista_capacity-resolution_2020, brunel_optimal_2004, brunel_is_2016}. 
Given a value of the load $\alpha$, condition (\ref{eq:const}) is satisfiable up to a maximum value of $k_{max}(\alpha)$. The maximum capacity achievable by the network is $\alpha_c^{P} = 2$, with $k_{max}(2) = 0$. Starting from any initialization of the coupling  matrix $J$ such that $J_{ii} = 0$ $\forall i$, all constraints in (\ref{eq:const}) will be satisfied after a number of iterations of the following serial update
\begin{align} 
    \label{eq:lp_update}
    \small
    &\delta J_{ij}^{(d)} = \lambda\sum_{\mu=1}^p\Theta\left(k-\Delta^\mu_i\right)\xi^\mu_i\xi^\mu_j, \hspace{0.3cm} \delta J_{ii}^{(d)}=0,
\end{align}
where $\lambda$ is a small positive learning rate, $d$ is the algorithm time step and $\Theta(x)$ is the Heaviside function.   
Perceptron learning is \textit{supervised}, in the sense that \cref{eq:lp_update} requires the explicit knowledge of the memories, at each step of training.
One can also symmetrize equation (\ref{eq:lp_update}) 
as 
\begin{equation}
    \begin{aligned}
    &\delta J_{ij}^{(d)} = \frac{\lambda}{2}\sum_{\mu=1}^p\Big[\Theta\left(k-\Delta^\mu_i\right) +\Theta\left(k-\Delta^\mu_j\right)\Big]\xi^\mu_i\xi^\mu_j, \\ 
    & \delta J_{ii}^{(d)}=0, 
    \end{aligned}
\end{equation}
obtaining a lower critical capacity and a general decreasing of the function $k_{max}(\alpha)$ due to the symmetry constraint \cite{benedetti_supervised_2022, gardner_phase_1989}. It has been shown numerically that, given $\alpha$, the larger is $k$, with $0 < k < k_{max}(\alpha)$, the wider are the basins of attraction. In line with the past literature \cite{battista_capacity-resolution_2020,scholkopf_learning_2018} we call a maximally stable perceptron, such that $k = k_{max}(\alpha)$, a Support Vector Machine (SVM). 

\subsection{Hebb's rule \label{sec:hebb}}
Hebb's (or Hebbian) learning prescription \cite{hopfield_neural_1982, hebb_organization_1949} consists in building up the connections between neurons as an empirical covariance of the memories, i.e. 
\begin{equation}
    \label{eq:hop}
    J_{ij}^{H} = \frac{1}{N}\sum_{\mu = 1}^p\xi_i^{\mu}\xi_j^{\mu}, \hspace{0.4cm}J_{ii}^{H} = 0.
\end{equation}
This rule allows retrieving memories up to a critical capacity $\alpha_c^H = 0.138$ \cite{amit_storing_1985}. Notably, when $\alpha < \alpha_c^H$ memories are not perfectly recalled, but only reproduced with a small number of errors. On the other hand, when $\alpha > \alpha_c^H$ interference between memories makes the learning rule completely ineffective, shifting the system into an oblivion regime. The landscape of attractors given by Hebb's rule is rugged and disseminated with \textit{spurious states}, i.e. stable fixed points of the dynamics that are far from the memories \cite{gardner_structure_1986}. 

\subsection{\label{sec:unl}Hebbian Unlearning}
Inspired by the brain functioning during REM sleep \cite{crick_function_1983}, the Hebbian unlearning algorithm (HU) \cite{hopfield_unlearning_1983, van_hemmen_increasing_1990,taylor_unlearning_1992,benedetti_supervised_2022, ventura_2024, benedettiEigenvectorDreaming2024} is a training procedure leading to classification and good generalization in a symmetric neural network. Training starts by initializing the connectivity matrix according to Hebb's rule \cref{eq:hop} (i.e. $J^{(0)} = J^H$). Then, the following procedure is iterated at each step $d$:
\begin{enumerate}
    \item Initialization of the network on a random neural state. 
    \item Run the asynchronous dynamics (\ref{eq:dynamics}) until convergence to a stable fixed point $\vec{S}^{(d)}$. 
    \item Update couplings according to:
    \begin{equation}
        \label{eq:unl_rule}
        \delta J_{ij}^{(d)} = -\frac{\lambda}{N}S_i^{(d)} S_j^{(d)},\hspace{0.3cm}\delta J_{ii}^{(d)} = 0,
    \end{equation}
\end{enumerate}
where $\lambda$ is a small positive learning rate. HU is a \textit{unsupervised} learning algorithm, in the sense that each step of training does not require explicit knowledge of the memories, but rather relies on implicit information encoded in the dynamics, i.e. in the current coupling matrix $J^{(d)}$. This algorithm was first introduced to prune the Hebbian energy landscape from proliferating spurious attractors \cite{gardner_structure_1986, amit_modeling_nodate}.
Classification is achieved by running the algorithm when $\alpha \leq \alpha_c^U$ with $\alpha_c^U \simeq 0.6$. In this regime, HU creates large basins of attraction around the memories, comparable in size to those of a SVM \cite{benedetti_supervised_2022}.

\section{\label{sec:twn}Training with noise}
The concept of learning from noisy examples, introduced in \cite{le_cun_learning_1986}, is at the basis of a pioneering study by Gardner and collaborators \cite{gardner_training_1989}, in an attempt to improve generalization in attractor neural networks. Here, we report the algorithm and characterize, for the first time, its performance over fully connected neural networks. Importantly, we find strong numerical evidence of a connection between the training-with-noise algorithm and the optimization problem studied in \cite{wong_optimally_1990,
wong_neural_1993}. The loss function for this problem, minimized by the training-with-noise algorithm, will be at the core of the analysis contained in our work. 
We will show that such an algorithm, in its original formulation, performs worse than a SVM. This outcome justifies the search for a new noise-injection prescription which will be explored in Section \ref{sec:The optimal noise condition}. 
\subsection{Gardner's algorithm}
The training-with-noise (TWN) algorithm \cite{gardner_training_1989} consists in starting from any initial coupling matrix $J^{(0)}$ with null entries on the diagonal, and updating recursively the couplings according to 
\begin{equation}
\label{eq:lwn}
    \small \delta J_{ij}^{(d)} = \frac{\lambda}{N}\Theta\left(-\xi_i^{\mu_d}\sum_{k=1}^N J_{ik}S_k^{\mu_d} \right) \xi_i^{\mu_d}S_j^{\mu_d}, \hspace{0.4cm} \delta J_{ii}^{(d)} = 0,
\end{equation}
where $\lambda$ is a small learning rate and $\mu_d \in [1,...,p]$ is a randomly chosen memory index. In this setting, $\vec{S}^{\mu_d}$ is a noisy data-point, generated as \begin{equation}
    \label{eq:chi}
    S_i^{\mu_d} = \chi_i^{\mu_d}\xi_i^{\mu_d},
\end{equation} with $\chi_i^{\mu_d}$ i.i.d. variables, sampled with probability
\begin{equation}
\label{eq:p_s}
    P(\chi_i^{\mu_d} = x) = \frac{(1+m_t)}{2}\delta(x - 1) + \frac{(1-m_t)}{2}\delta(x + 1).
\end{equation}
Notice that $\vec{S}^{\mu_d}$ are binary variables, as the $\xi_i^{\mu_d}$ are. The $\textit{training overlap}$ $m_t$ is a control parameter for the level of \textit{noise} injected during training, corresponding to the expected overlap between $\vec{S}^{\mu_d}$ and $\vec{\xi}^{\mu_d}$, i.e.
\begin{equation}
    \label{eq:mt}
    m_t = \frac{1}{N}\sum_{i=1}^N\xi_i^{\mu_d} S_i^{\mu_d} + O\Big(\frac{1}{\sqrt{N}}\Big).
\end{equation}
The algorithm would converge when every configuration with overlap $m_t$ with a memory generates on each site a local field aligned with the memory itself.
TWN is a \textit{supervised} learning algorithm, in the sense that \cref{eq:lwn} requires the explicit knowledge of the memories, at each step of training.\\

\subsection{A loss function for TWN}
The goal of this subsection is to connect the purely empirical TWN introduced in \cite{gardner_training_1989} with the optimization problem studied in \cite{wong_neural_1993, wong_optimally_1990}. We show that TWN globally minimizes a loss function. This will allow us to get valuable analytical insights on its performance. \\  
Let us define the function
\begin{equation}
    \label{eq:loss_ws}
    \mathcal{L}(m, J) = -\frac{1}{\alpha N^2}\sum_{i,\mu}^{N,p} \text{erf} \left(\frac{m\Delta_i^{\mu}}{\sqrt{2(1-m^2)}}\right).
\end{equation}
Wong and Sherrington \cite{wong_optimally_1990,
wong_neural_1993} propose an elegant analysis of a  network designed to optimize $\mathcal{L}(m, J)$ given $m$, i.e. whose couplings correspond to the global minimum of $\mathcal{L}(m, J)$. The physical meaning of $\mathcal{L}(m, J)$ comes from the identity 
\begin{equation}
\label{eq:onestep_ret_map}
\small
    \mathcal{L}(m,J) = - \overline{\frac{1}{N}\sum_{i = 1}^N  \Big\langle \xi_i^\nu \;\text{sign}\Big(\sum_{j=1}^N J_{ij}S_j^\nu(0)\Big)\Big\rangle}\;,
\end{equation}
where $S^\nu(0)$ is a configuration with overlap $m$ with memory $\xi^\nu$. The RHS of \cref{eq:onestep_ret_map} is the opposite of the \textit{one-step retrieval map}, defined by selecting one memory, sampling a noisy state having overlap $m$ with the memory, and measuring the overlap obtained after a single step of \textit{synchronous} dynamics (\ref{eq:dynamics}). Hence, minimizing $\mathcal{L}(m, J)$ means driving the first step of the retrieval dynamics towards the memories, enlarging their attraction basins. \\
The coupling matrix $J(m)$ resulting from the optimization of \cref{eq:loss_ws} depends on $m$.  Some of their findings, relevant to this work, are:
\begin{enumerate}
    \item For any $m_0$, the maximum value of the one-step retrieval map $-\mathcal{L}(m_0, J(m))$ is obtained at $m = m_0$.
    \item When $m \rightarrow 1^-$, the minimization of the function $\mathcal{L}(m,J(m))$ trains a linear perceptron with maximal stability, i.e. a SVM. 
    \item When $m \rightarrow 0^+$, the  minimization of $\mathcal{L}(m,J(m))$ leads to a Hebbian connectivity matrix $J_{ij} \propto J_{ij}^{H}$.
\end{enumerate}

We now proceed to establish a connection between the TWN procedure and the theoretical results obtained by Wong and Sherrington. When the network is trained through TWN, the resulting coupling matrix depends on $m_t$, i.e. $J = J(m_t)$. 
It is crucial to stress the difference between the variables $m$ and $m_t$: the former is a parameter of eq.~(\ref{eq:loss_ws}), the latter is the level of noise used by the training algorithm. 
It comes out that eq. (\ref{eq:loss_ws}) is relevant to the TWN procedure, since iterating \cref{eq:lwn} leads to a decrease of $\mathcal{L}(m, J(m_t))$, for any value of $m$ and $m_t$. In fact, considering a small variation of the stabilities induced by the algorithm update
\begin{equation*}
     \Delta_i^{\mu} \rightarrow \Delta_i^{\mu} + \delta \Delta_i^{\mu},
\end{equation*}
and performing a Taylor expansion of (\ref{eq:loss_ws}) at first order in $O(N^{-1/2})$, one obtains (see Appendix \ref{sec:appA1})
\begin{equation}
    \label{eq:new_loss3}
    \mathcal{L}^{'} = \mathcal{L} + \sum_{i=1}^{N}
    \delta\mathcal{L}_{i},
\end{equation}
where 
\begin{equation}
\begin{aligned} 
\label{eq:delta_loss}
\small
    \delta\mathcal{L}_{i} = -&\Theta\left(-\xi_i^{\mu_d}\sum_{k=1}^N J_{ik}S_k^{\mu_d} \right)\frac{\lambda}{\alpha\sigma_i N^{5/2}}\cdot\\
    \cdot &\frac{\sqrt{2}m \cdot m_t}{\sqrt{\pi(1-m^2)}}\exp{\left(-\frac{m^2 \Delta_i^{\mu_d^2}}{2(1-m^2)}\right)}.
    \end{aligned}
\end{equation}
Hence, $\delta\mathcal
{L}_{i}$ is strictly non-positive when $\frac{\lambda}{N}$ is small, so that the Taylor expansion is justified. 
\begin{figure}[ht!]
\centering
\includegraphics[width=0.9\linewidth]{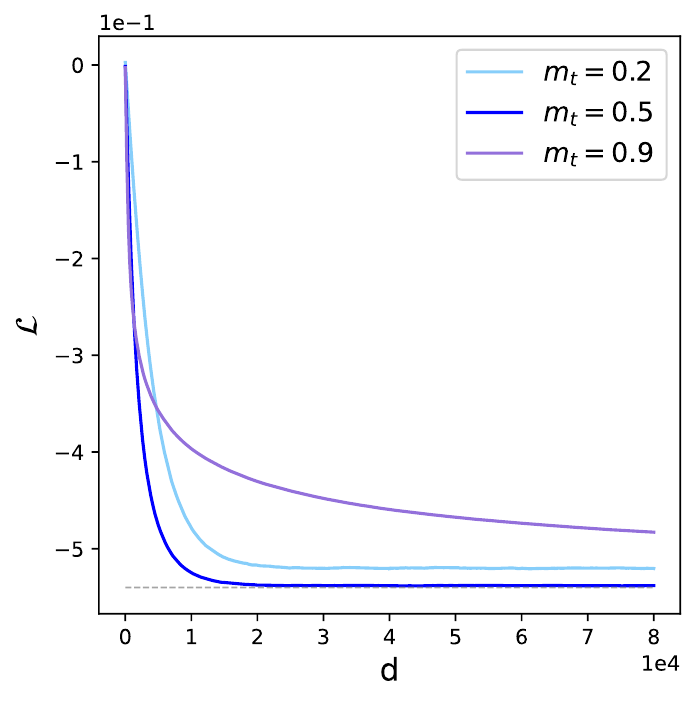}
\caption{Lines in the main plot report the function $\mathcal{L}(m = 0.5, J(m_t))$ for different training overlaps as functions of the number of algorithm steps $d$. The \textit{dashed} line represents the theoretical minimum value from \cite{wong_neural_1993}. All measures are averaged over $5$ realizations of the couplings $J$. Choice of the parameters: $N = 100$, $\alpha = 0.3$, $\lambda = 1$, the initial couplings are Gaussian with unitary mean, zero variance and $J_{ii}^{(0)} = 0$ $\forall i$.}
\label{fig:loss}
\end{figure}
Moreover, we numerically find that iterating (\ref{eq:lwn}) with a given value of $m_t$ drives $\mathcal{L}(m=m_t, J(m_t))$ to its theoretical absolute minimum computed in \cite{wong_neural_1993}, as reported in \cref{fig:loss}. As a consequence, the performance of the TWN algorithm can be completely described in the analytical framework of \cite{wong_neural_1993} and $\mathcal{L}(m = m_t, J)$ can be considered as the loss function optimized by the TWN algorithm. \\

\subsection{TWN performance}
The same analytical techniques applied in \cite{wong_neural_1993} can be used to determine whether the TWN algorithm is capable to reach perfect classification. We find that the distribution of stabilities in the trained network (see equation (\ref{eq:rho})) has always a tail in the negative values, implying that classification is never reached. The only exception to this statement are the trivial cases of $m_t = 1^-$ (SVM limit) and $m_t = 1$ (linear perceptron with $k = 0$), where all memories are stable for $\alpha \leq 2$. \\

\begin{figure}[t]
\centering
\includegraphics[width=0.95\linewidth]{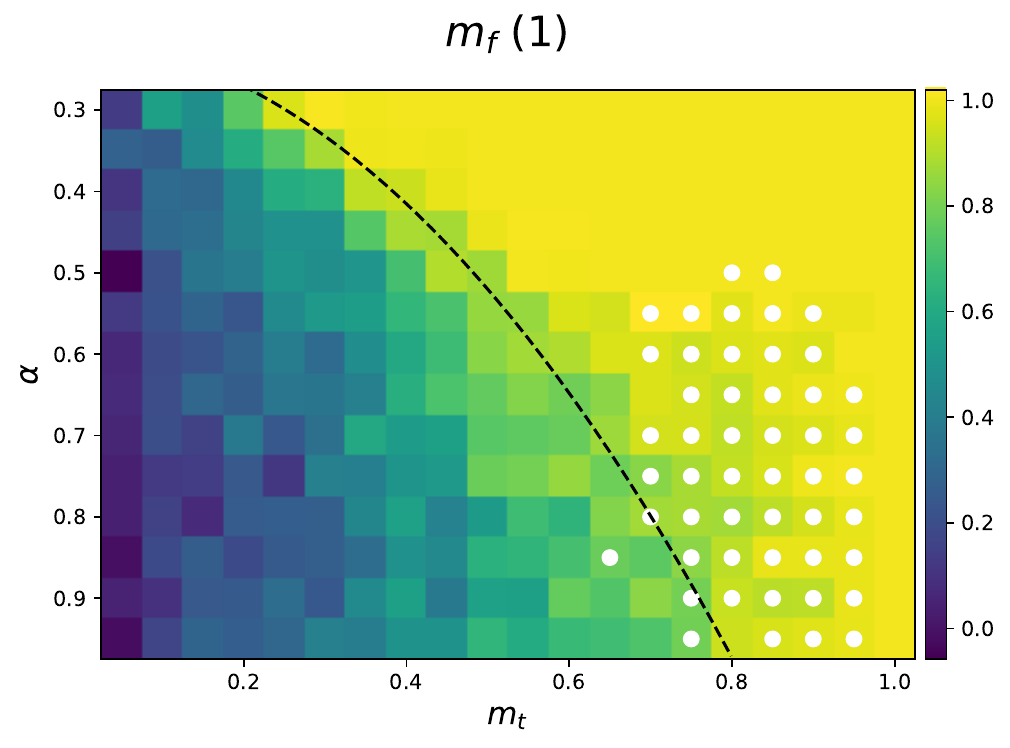}
\caption{$m_f(1)$ as a function of $m_t$ and $\alpha$. Warmer shades of color are associated to higher retrieval performances. The \textit{black dashed} line represents the boundary of the retrieval regime according to the criterion in Appendix \ref{app:c}, \textit{white} dots signal the points where basins of attraction to which memories belong are larger than ones obtained from a SVM at $N = 200$.} 
\label{fig:mf1}
\end{figure}
The generalization properties of a network trained through TWN are now discussed. The color map in \cref{fig:mf1} reports the numerical estimate of the retrieval map $m_f$ at $m_0 = 1$, extrapolated in the $N\rightarrow \infty$ limit, as a measure of the distance between a given memory and the closest attractor.  
In Appendix \ref{app:c} we propose an empirical criterion to separate a \textit{retrieval} phase from a \textit{non-retrieval} one. In the first region, the memory appears to close to an attractor when $N\rightarrow \infty$, while in the second region the closest attractor is typically orthogonal in the thermodynamic limit.
Such criterion is based on assuming that $m_f(m_0)$, measured with respect to the attractor of the basin to which a given memory belongs and not the memory itself, always develops a plateau starting in $m_0 = 1$ and ending in some $m_c < 1$ when $N \rightarrow \infty$. The behavior of the basin radius is then observed numerically as a function of $m_t$: when the plateau disappears (i.e. $m_c = 1$) then one can suppose that basins get shattered in the configurations space due to the interference with the other attractors, and this occurs at some value of $m_t$. The transition between the two regions is reported as a dashed line in \cref{fig:mf1}.\\ 
In the retrieval region, we compute the typical size of the basins of attraction, as described in Appendix \ref{app:c}. White dots in \cref{fig:mf1} signal the combinations of $(m_t, \alpha)$ where the basins obtained by TWN algorithm resulted larger than the ones shaped by a SVM at the same value of $\alpha$. One can conclude that for most of the retrieval region the generalization performance is worse than the SVM, which maintains larger basins of attraction; on the other hand, at higher values of $\alpha$ the trained-with-noise network sacrifices its classification property to achieve a basin that appears wider than the SVM one. We want to stress the importance of a comparison between the TWN and the corresponding SVM, since numerical investigations have shown the latter to achieve extremely large basins of attraction, presumably due to the maximization of the stabilities \cite{forrest_content-addressability_1988, benedetti_supervised_2022}.\\

In conclusion, the TWN algorithm never outperforms the corresponding SVM in terms of attraction basins size without reducing its classification capabilities. This observation moves our research for a modification to the TWN algorithm, which can ensure a performance that is similar to the SVM one even at lower values of $m_t$ and higher loads $\alpha$.   

\section{The optimal structure of noise \label{sec:choice_of_training_conf}}

As previously stated, SVMs are considered to be highly efficient associative memory models, due to their very good classification and generalization capabilities. Wong and Sherrington's analytical argument proves that minimization of $\mathcal{L}(m=1^-, J)$ trains a SVM. 
The TWN procedure proposed by Gardner and collaborators, relying on a Bernoulli process to generate noise, accomplishes this task only when $m_t = 1^-$ (see Section~\ref{sec:twn}). Injecting a larger amount of noise during training would deteriorate the performance: specifically, $m_t = 0^+$ trains a Hebbian neural network. 
Nevertheless, training a network with examples that are nearly uncorrelated with the memories can significantly speed up the sampling process, since such states can be generated in an unsupervised fashion, without knowledge of the memories (as seen for the HU algorithm in Section~\ref{sec:unl}). Such unsupervised processes are also considered more biologically plausible. \\ 
In this Section, we show that it is possible to use maximally noisy configurations (i.e. $m_t = 0^+$) to train a network that resembles a SVM, by means of the TWN algorithm. This translates into approaching the global minimum of $\mathcal{L}(m = 1^-, J(0^+))$. For this purpose, one must change the way noisy data are generated: they need to meet specific constraints which lead to internal dependencies among the features (i.e. a \textit{structure}). We derive a theoretical condition characterizing the optimal structure of noise, and show that specific configurations in the Hebbian energy landscape, including local minima, match well the theoretical requirements.\\
\subsection{The optimal noise condition}
\label{sec:The optimal noise condition}
It will be helpful for our purposes to implement a symmetric version of rule (\ref{eq:lwn}), i.e.
\begin{equation}
    \begin{aligned}
    \label{eq:lwn2}
    \delta J_{ij}^{(d)} = \frac{\lambda}{N}\Big(&\Theta\big(-\xi_i^{\mu_d}\sum_{k=1}^N J_{ik}S_k^{\mu_d}\big) \xi_i^{\mu_d} S_j^{\mu_d} + \\ 
    + &\Theta\big(-\xi_j^{\mu_d}\sum_{k=1}^N J_{jk}S_k^{\mu_d} \big) \xi_j^{\mu_d} S_i^{\mu_d}\Big).
    \end{aligned}
\end{equation}
Henceforth in the paper, TWN will be understood as the symmetric update rule \cref{eq:lwn2}. Since SVMs show a high degree of symmetry in the J matrix \cite{battista_capacity-resolution_2020, brunel_is_2016}, we extend the results presented in Section \ref{sec:twn} to the current case. This assumption is well-supported by further numerical analysis. Equation (\ref{eq:lwn2}) can be reformulated explicitly by rewriting the Heaviside function as $\Theta(x)=\big(1+\text{sign}(x)\big)/2$, leading to
\begin{equation}
    \begin{aligned}
    \label{eq:lwsn}
    \delta J_{ij}^{(d)} = &\frac{\lambda}{2N}\big(\xi_i^{\mu_d}S_j^{\mu_d} + S_i^{\mu_d}\xi_j^{\mu_d}\big) \\ -&\frac{\lambda}{2N}\big(S_i^{1,\mu_d}S_j^{\mu_d} + S_i^{\mu_d}S_j^{1,\mu_d}\big),
    \end{aligned}
\end{equation}
where $S_i^{1,\mu_d} := \text{sign}\big(\sum_{k=1}^N J_{ik} S_k^{\mu_d}\big)$ corresponds to one step of the dynamics. In the maximal noise case $m_t=0^+$, the variation of $\mathcal{L}(m, J)$ in one step of the algorithm is then (see Appendix~\ref{sec:appA2})
\begin{equation}
    \label{eq:deltaL_lwsn}
    \small
    \delta\mathcal{L} \propto \frac{m}{\sqrt{2\pi(1-m^2)}}\sum_{i,\mu}^{N,p}\omega_i^{\mu}\exp{\left(-\frac{m^2\Delta_i^{\mu^2}}{2(1-m^2)}\right)}, 
\end{equation}
where the weights $\omega_i^{\mu}$ are given by
\begin{equation}
    \label{eq:omegatilde}
    \omega_i^{\mu} =\frac{1}{2\sigma_i}\left( m_{\mu}\chi_i^{1,\mu} + m_{1,\mu}\chi_i^{\mu}\right),
\end{equation}
with 
\begin{equation}
\label{eq:def_chi}
    \chi_i^{\mu} = \xi_i^{\mu}S_i^{\mu_d},\hspace{1cm}\chi_i^{1,\mu} = \xi_i^{\mu}S_i^{1,\mu_d},
\end{equation}
and
\begin{equation}
\label{eq:def_m}
    m_{\mu} = \frac{1}{N}\sum_{j = 1}^N S_j^{\mu_d}\xi_j^{\mu},\hspace{1cm}m_{1,\mu} = \frac{1}{N}\sum_{j = 1}^N S_j^{1,\mu_d}\xi_j^{\mu}.
\end{equation}
Notice that definitions \cref{eq:def_chi} and \cref{eq:def_m} mirror \cref{eq:chi} and \cref{eq:mt}, but describe the relation between training data points and a \textit{generic} memory $\vec{\xi}^{\mu}$, not just the specific $\vec{\xi}^{\mu_d}$ picked for the training step. Since our goal is to approach the performance of a SVM, we are interested in the limit $m\rightarrow 1^-$.  
When $m\rightarrow 1^-$, the Gaussian factors in \cref{eq:deltaL_lwsn} are tightly peaked around zero. Since we want $\delta \mathcal{L}$ to be negative, we need, for most of the pairs $i,\mu$,
\begin{equation}
    \label{eq:deltaL_lwsn3}
    \omega_i^{\mu} < 0 \hspace{0.2cm}\text{if}\hspace{0.2cm}\Delta_i^\mu\sim0.
\end{equation}
\\The more negative $\omega_i^{\mu}$ is when $\Delta_i^\mu \sim 0$, the more powerful is its contribution to approach the SVM performances. Equation (\ref{eq:deltaL_lwsn3}) is the main result of our work: it gives the condition that maximally noisy states $\vec{S}^{\mu_d}$ must satisfy to train an attractor neural network approaching the performance of a SVM. Selecting training data that satisfy equation (\ref{eq:deltaL_lwsn3}) amounts to imposing specific internal dependencies among the noise units $\vec{\chi}$, which are no more i.i.d. random variables, as it was in \cite{gardner_training_1989}. We refer to such dependencies as \textit{structure} of the 
 noise.
 One should also bear in mind that training is a dynamic process: to reduce $\mathcal{L}(m=1^-, J(0^+))$  condition (\ref{eq:deltaL_lwsn3}) should hold throughout training.

\subsection{Numerical Analysis} 
\begin{figure*}[ht!]
     \centering
     \begin{subfigure}[b]{0.43\textwidth}
         \centering
         \includegraphics[width=\textwidth]{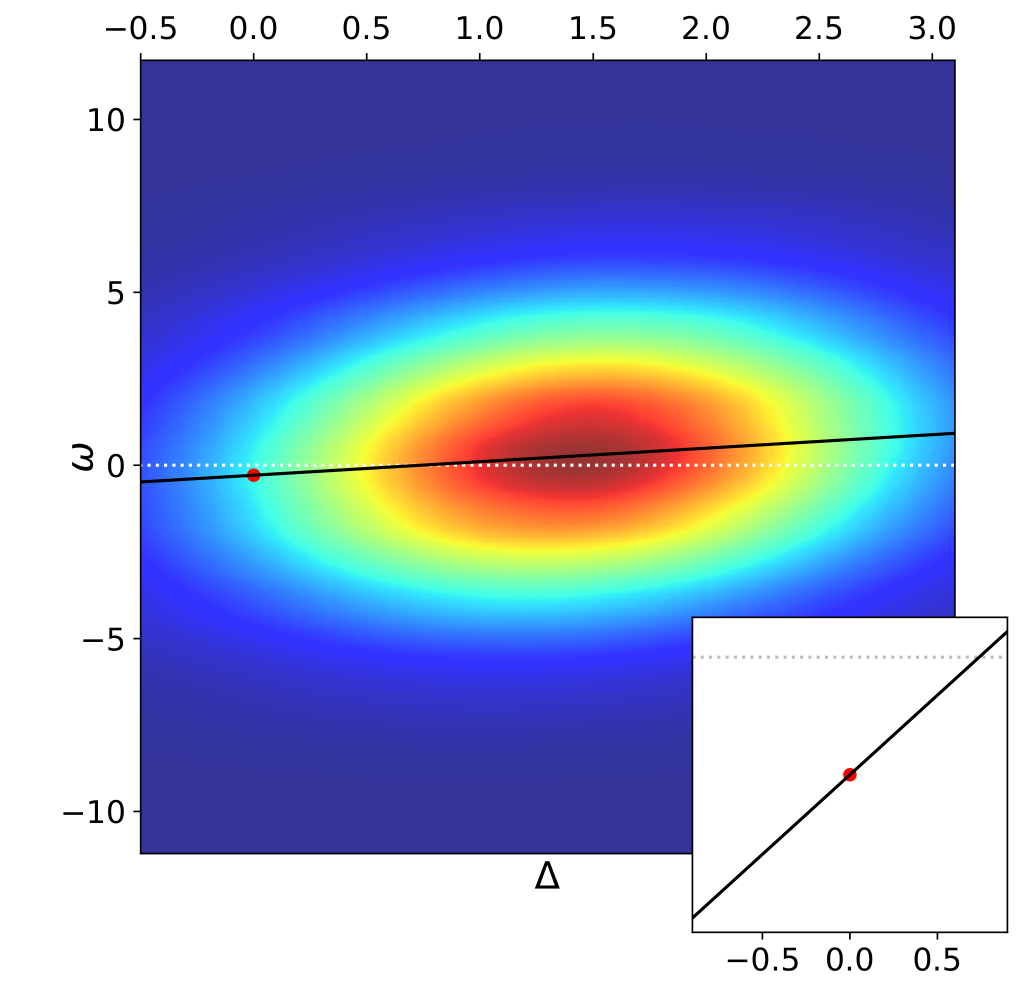}
         \caption{}
     \end{subfigure}
     \hfill
     \begin{subfigure}[b]{0.4\textwidth}
         \centering
         \includegraphics[width=\textwidth]{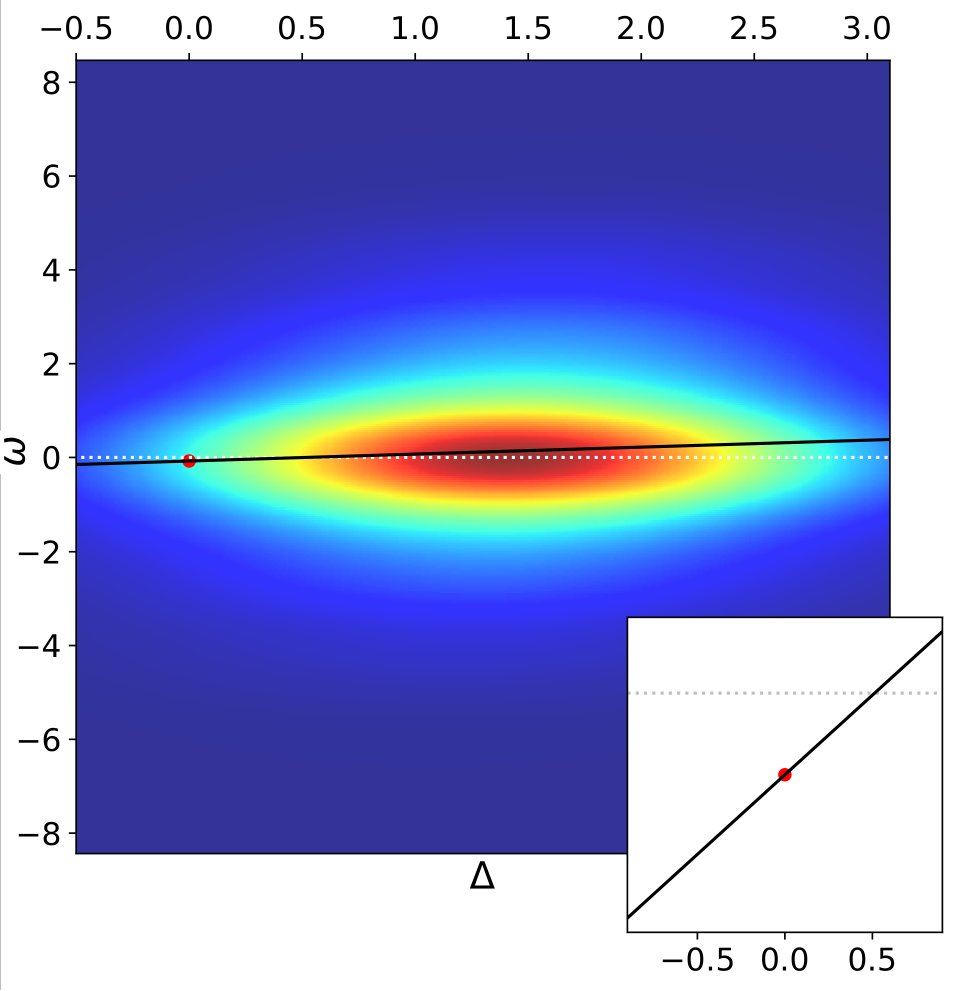}
         \caption{}
     \end{subfigure}
          \begin{subfigure}[b]{0.4\textwidth}
         \centering
         \includegraphics[width=\textwidth]{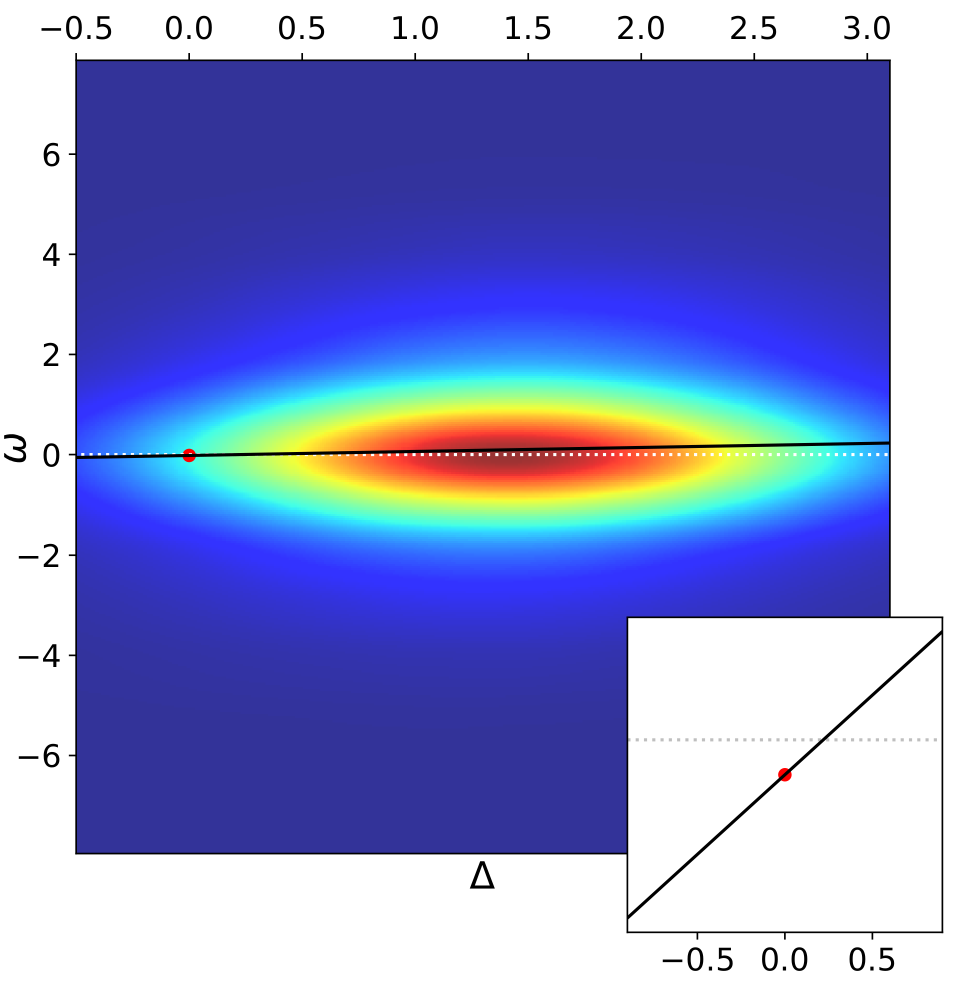}
         \caption{}
     \end{subfigure}
     \hfill
     \begin{subfigure}[b]{0.4\textwidth}
         \centering
         \includegraphics[width=\textwidth]{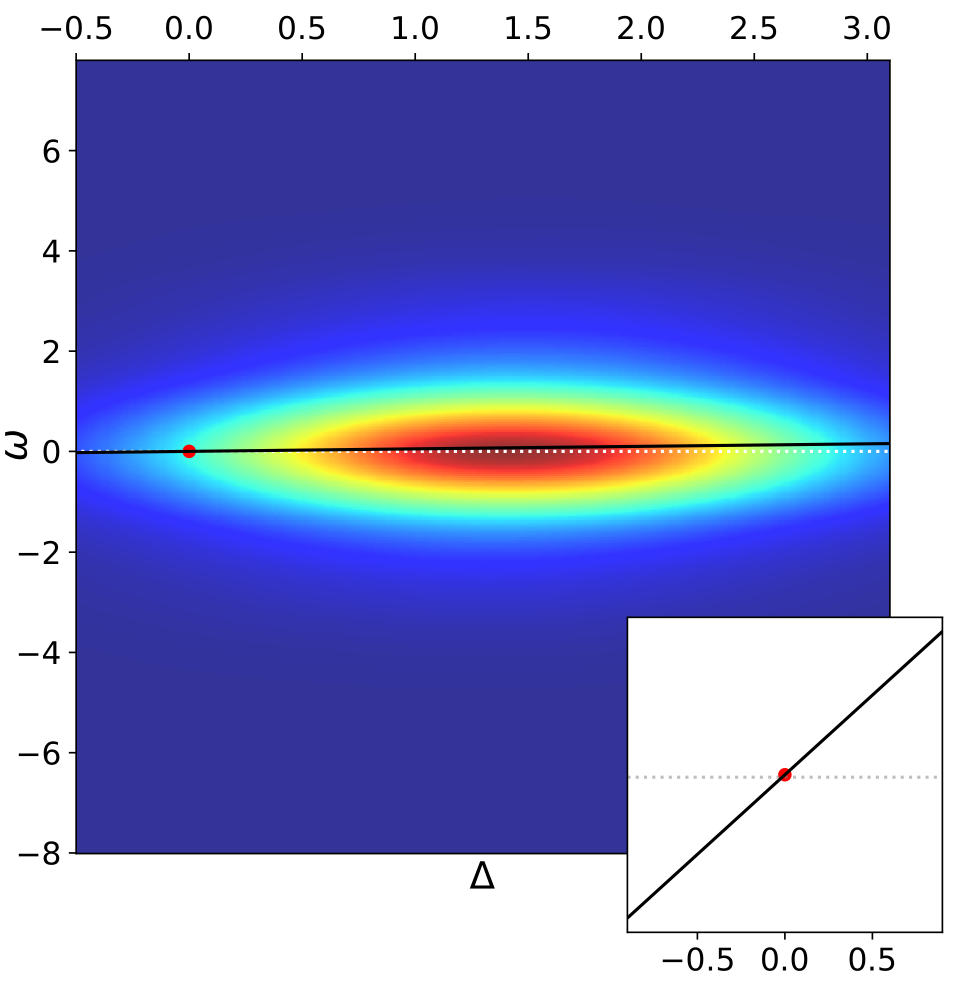}
         \caption{}
     \end{subfigure}
     \hfill
        \caption{Distribution of $\omega_i^{\mu}$ as a function of $\Delta_i^{\mu}$ for training configurations sampled with a Monte Carlo at temperature $T = 0$ i.e. stable fixed points only (a), $T = 0.5$ (b), $T = 1$ (c), $T = 8$ (d), on a Hebbian network. Warmer colors represent denser region of data points. The \textit{full black} line is the non-weighted best fit line for the points, the \textit{dotted white} line represents $\omega = 0$, the \textit{red dot} is the value of the best fit line associated with $\Delta = 0$. Sub-panels to each panel report a zoom of the line around $\Delta = 0$. Measures have been collected over $15$ samples of the network, and observations show that finite size effects are negligible. Choice of the parameters: $N = 500$, $\alpha = 0.5$.}
        \label{fig:omegatilde}
\end{figure*}
\begin{figure*}[ht!]
     \centering
     \begin{subfigure}[b]{0.42\textwidth}
         \centering
         \includegraphics[width=\textwidth]{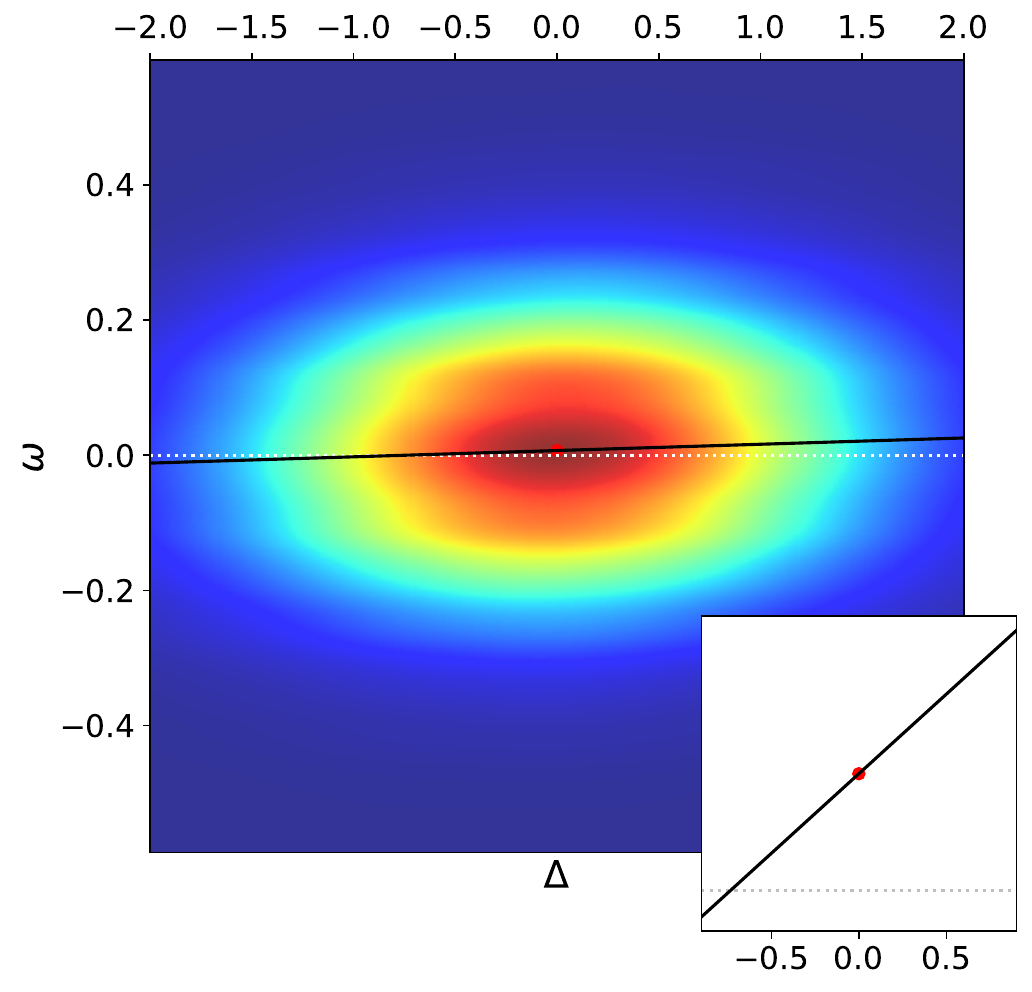}
         \caption{}
     \end{subfigure}
     \hfill
     \begin{subfigure}[b]{0.4\textwidth}
         \centering
         \includegraphics[width=\textwidth]{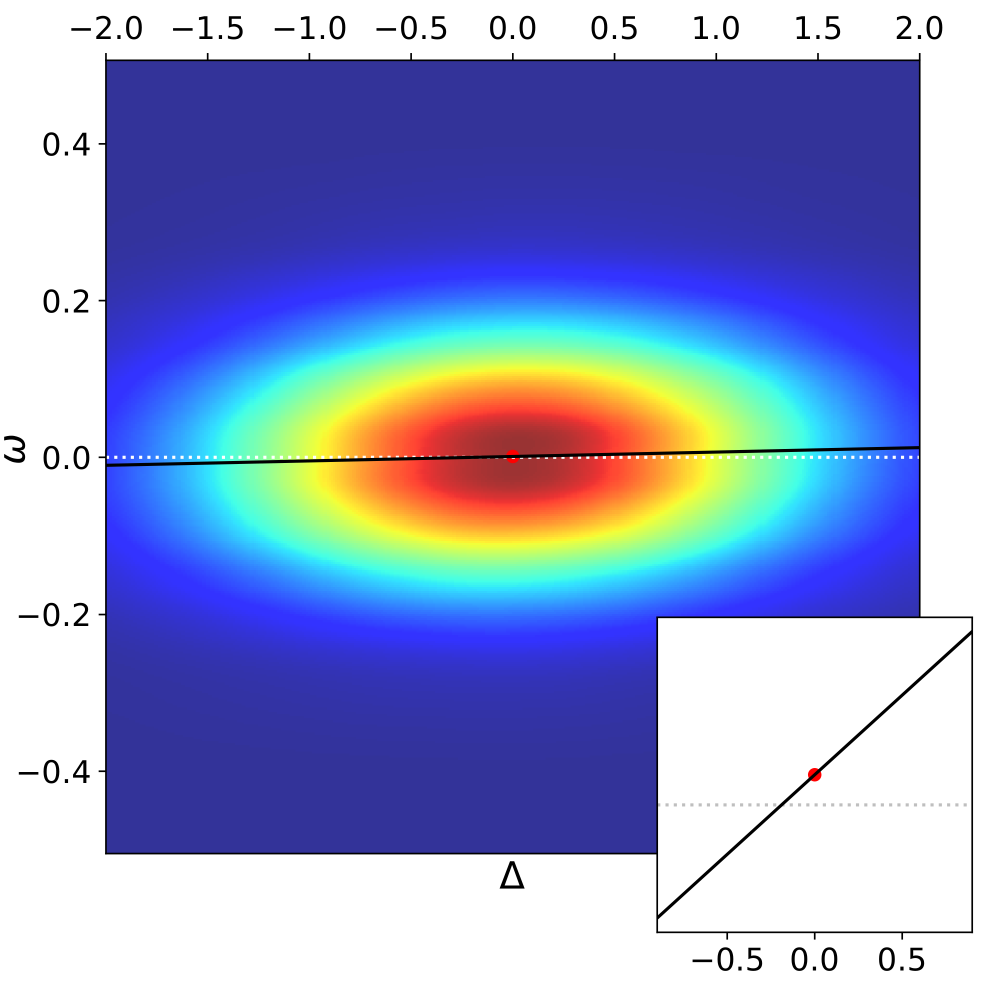}
         \caption{}
     \end{subfigure}
          \begin{subfigure}[b]{0.42\textwidth}
         \centering
         \includegraphics[width=\textwidth]{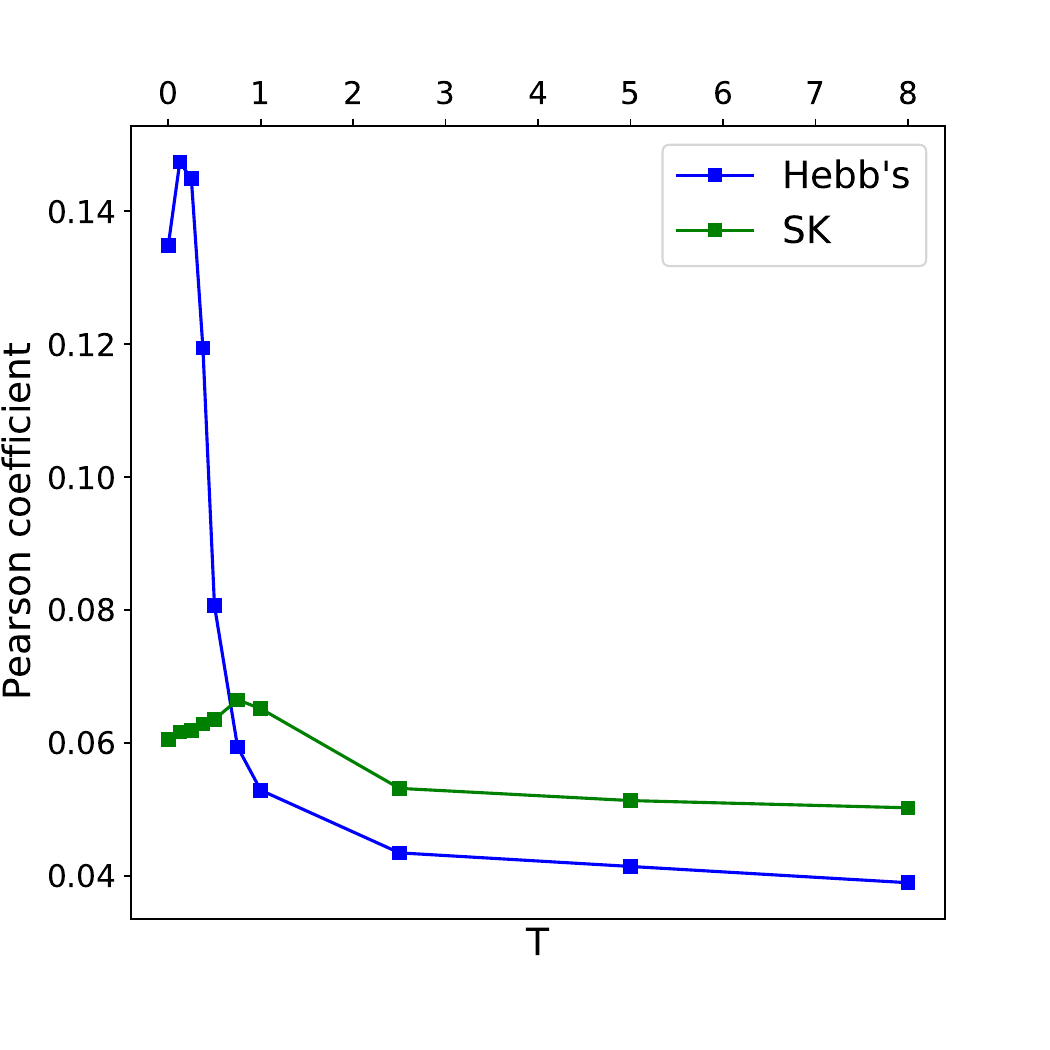}
         \caption{}
     \end{subfigure}
     \hfill
     \begin{subfigure}[b]{0.42\textwidth}
         \centering
         \includegraphics[width=\textwidth]{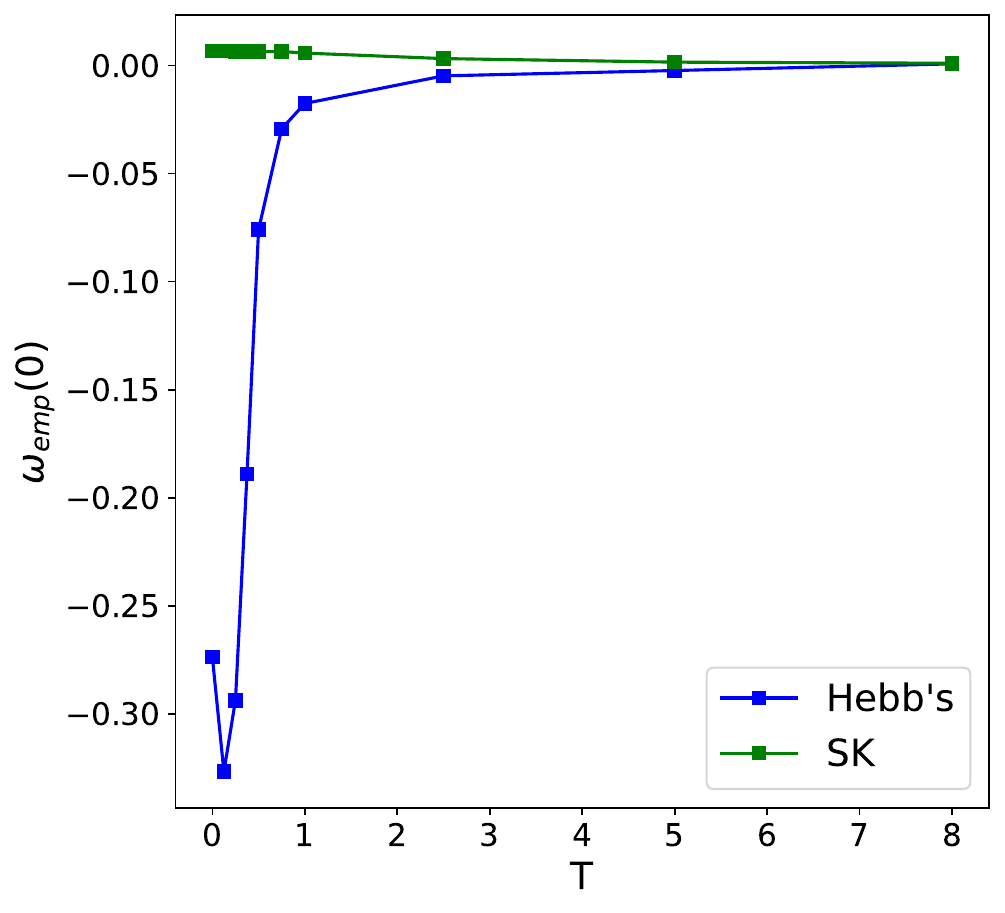}
         \caption{}
     \end{subfigure}
     
     \hfill
        \caption{(a), (b): Distribution of $\omega_i^{\mu}$ as a function of $\Delta_i^{\mu}$ for training configurations sampled with a Monte Carlo at temperature $T = 0$ i.e. stable fixed points only (a), and $T = 8$ (b) on a SK model. Warmer colors represent denser region of data points. The \textit{full black} line is the non-weighted best fit line for the points, the \textit{dotted white} line represents $\omega = 0$, the \textit{red spot} is the value of the best fit line associated with $\Delta = 0$. Sub-panels to each panel report a zoom of the line around $\Delta = 0$. (c), (d): Comparison between the Hebbian initialization and the Random one through evaluation of: the Pearson coefficient between $\omega_i^{\mu}$ and $\Delta_i^{\mu}$ (c) and the estimated value of $\omega_{emp}(0)$ from the dispersion plots (d). Measures have been collected over $15$ samples of the network. Choice of the parameters: $N = 500$, $\alpha = 0.5$.}
        \label{fig:omegatilde_SK}
\end{figure*}
\begin{figure*}[ht!]
\begin{tabularx}{\linewidth}{*{3}{X}}
     \begin{subfigure}[b]{\linewidth}
         \centering
         \includegraphics[width=0.96\textwidth]{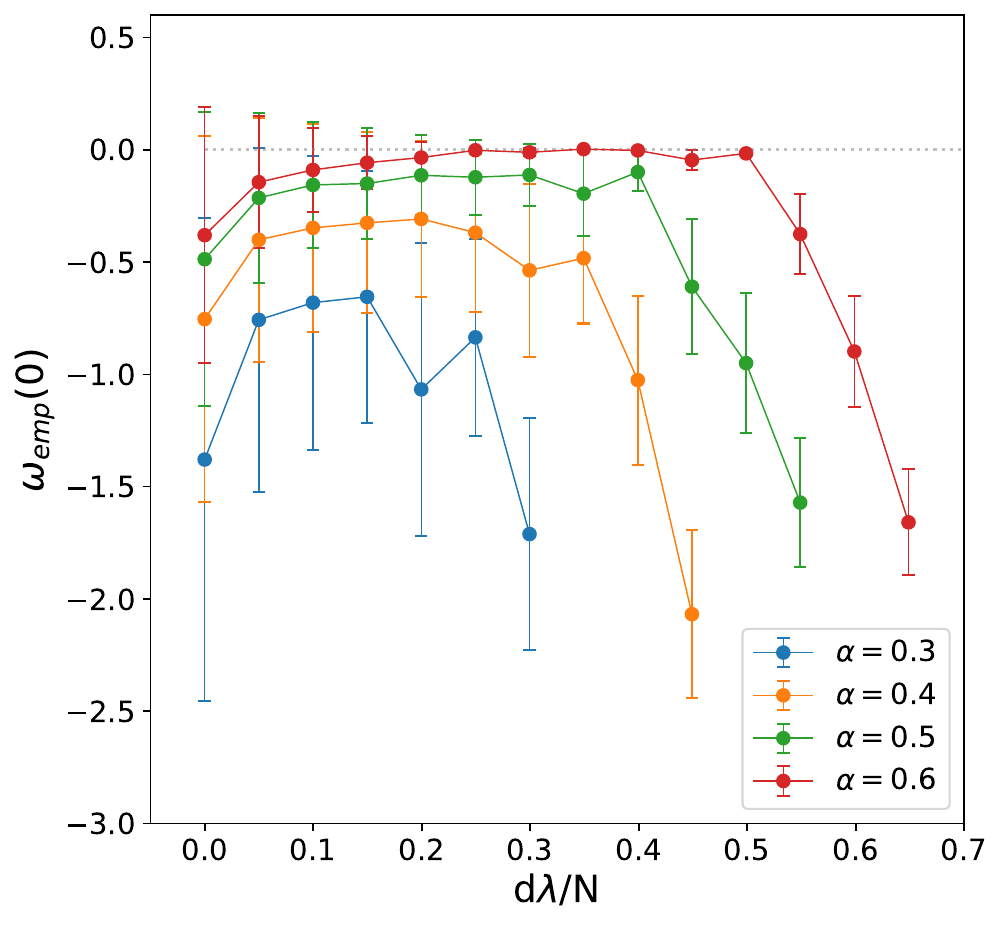}
         \caption{}
     \end{subfigure}
     &
     \hfill
     \begin{subfigure}[b]{\linewidth}
         \centering
         \includegraphics[width=\textwidth]{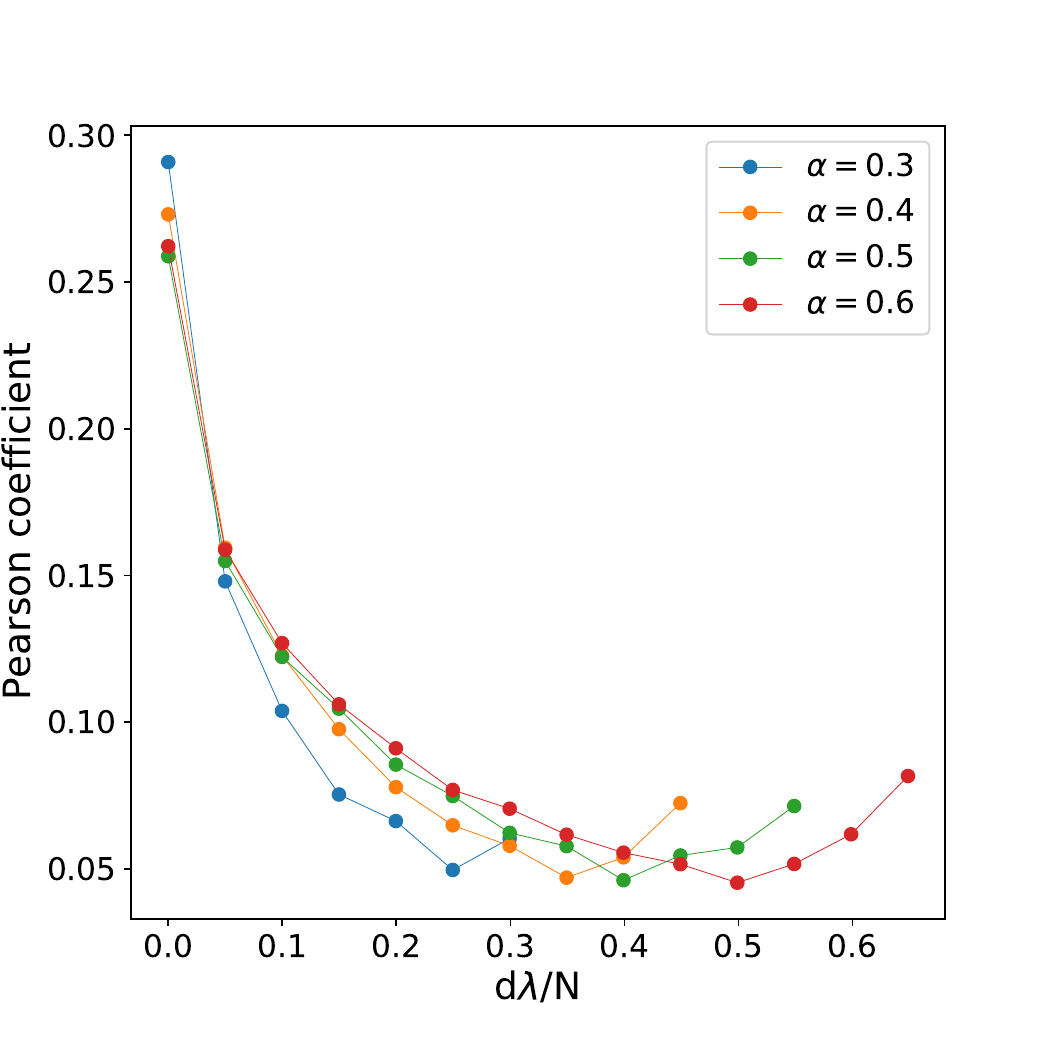}
         \caption{}
     \end{subfigure}
     &
     \hfill
     \begin{subfigure}[b]{\linewidth}
         \centering
         \includegraphics[width=0.93\textwidth]{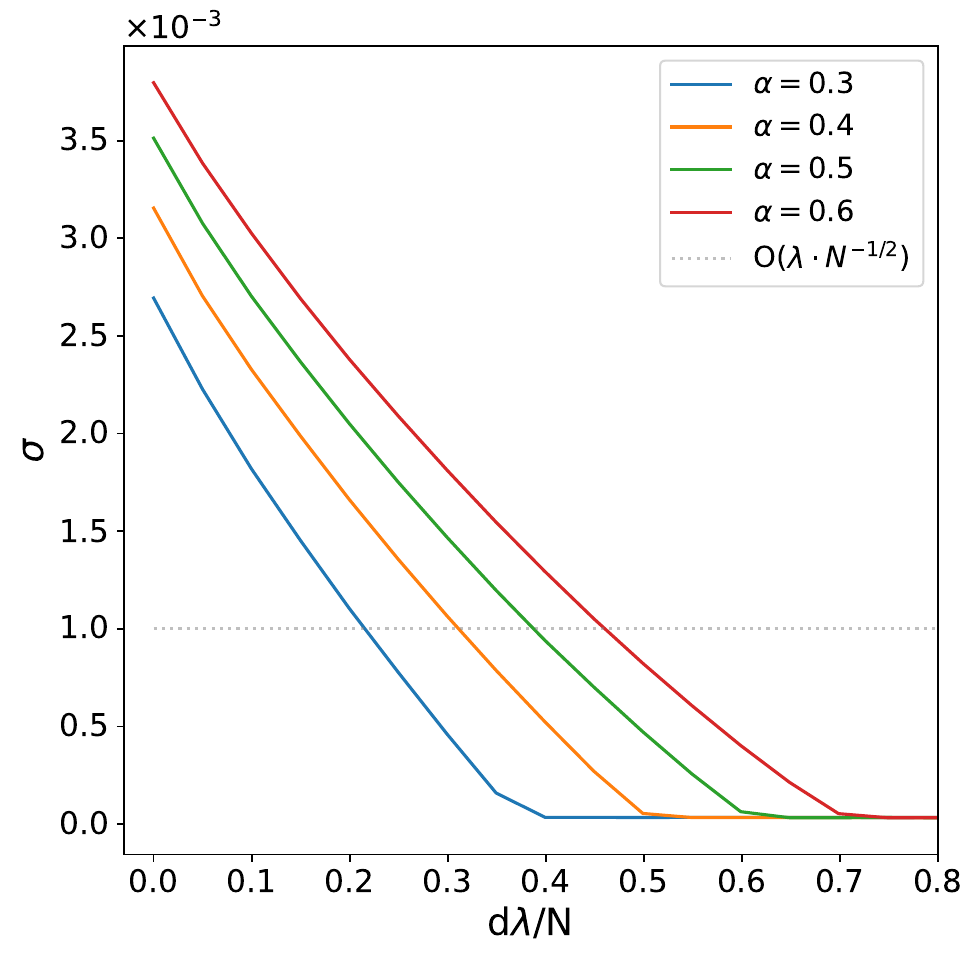}
         \caption{}
     \end{subfigure}
\end{tabularx}
\caption{The TWN algorithm is implemented by sampling stable fixed points of the network dynamics with $m_t = 0^+$. (a) The empirical measure of $\omega_i^{\mu}$ around $\Delta_i^\mu = 0$ for the case of stable fixed points as a function of the rescaled number of iterations of the learning algorithm. Error bars are given by the standard deviations of the measures. (b) Pearson coefficient measured between $\omega_i^{\mu}$ and $\Delta_i^{\mu}$. (c) The standard deviation of the couplings during learning, defined as $\sigma = \frac{1}{N}\sum\sigma_i$. Points are averaged over $50$ samples and the choice of the parameters is: $N = 100$, $\lambda = 10^{-2}$.}
\label{fig:omega_dynamic}     
\end{figure*}
As seen in section \ref{sec:twn}, traditional TWN relies on fully random states, i.e. very high states in the energy landscape defined by the function 
\begin{equation}
    E[\vec{S}|J] = -\frac{1}{2}\sum_{i,j}S_i J_{ij}S_j.
\end{equation} 
Since $E$ plays the role of the Lyapunov function of the dynamics when couplings $J_{ij}$ are symmetric \cite{amit_modeling_nodate}, the dynamic rule in \cref{eq:dynamics} always converges to local minima of $E$, in the same framework. Inspired by other learning algorithms that make use of particular states in the energy landscape, e.g. the HU procedure described in section \ref{sec:unl}, we want to test numerically whether configurations sampled at different altitudes of the landscape are characterized by an internal structure that satisfies condition (\ref{eq:deltaL_lwsn3}).    
To do so, we sample training configurations according to the Gibbs-Boltzmann measure by means of a Monte Carlo routine at temperature $T$. Temperature acts as a control parameter: when $T=0$ training configurations are stable fixed points of \cref{eq:dynamics}. Higher values of $T$ progressively reduce the structure of noise in training configurations, and in the limit $T\to\infty$, training configurations are the same as in the TWN algorithm. The Monte Carlo of our choice is of the Kawasaki kind \cite{newman_monte_1999}, to ensure that all training configurations are at overlap\footnote{For the sake of the experiment, we say that a state has overlap $m_t = 0^+$ with one memory $\vec{\xi}$ if $m_t\in (0,N^{-1/2})$.} $m_t = 0^+$. We are going to use this technique to probe the states across two types of landscapes: the one resulting from a Hebbian initialization and the one resulting from a Sherrington-Kirkpatrick (SK) model \cite{sherrington_solvable_1975}, where couplings are i.i.d. Gaussian with zero mean. \\

Figure~\ref{fig:omegatilde} reports numerical results for a network initialized according to Hebb's prescription \cref{eq:hop}, for four different temperatures. Each panel shows the distribution of $(\Delta_i^\mu, \omega_i^{\mu})$. Data points are collected over fifteen realizations of the network, then plotted and smoothed to create a density map. We are interested in the \textit{typical} behavior of $\omega_i^{\mu}$ when $\Delta_i^\mu=0$, which can be estimated by a linear fit of the data. We consider the intercept $\omega_{emp}(0)$ of the best fit line as an \textit{indicator} of the typical value of $\omega_i^{\mu}$ around $\Delta_i^{\mu} = 0$. 
We find that at high temperature $\omega_{emp}(0)$ gets closer to zero, suggesting low quality in terms of training performance. On the other hand, at lower temperatures the sampled training configurations favor both classification and generalization, since $\omega_{emp}(0)$ is more negative. \\
For a comparison, one can study the distribution of $\omega_i^{\mu}$ in the case of a random initialization of the coupling matrix $J$. We chose the SK model \cite{sherrington_solvable_1975} as a case of study. 
Panels (a) and (b) in \cref{fig:omegatilde_SK} report the smoothed distribution of $(\Delta_i^\mu, \omega_i^{\mu})$  showing a different scenario with respect to the Hebbian one. The distribution looks anisotropic, as in the Hebbian case, yet the stabilities are Gaussians centered around zero and $\omega_{emp}(0)$ is positive. In particular, things appear to improve when $T$ increases, in contrast with the previous case of study.  
In order to quantify the correlation between $\omega_i^{\mu}$ and $\Delta_i^\mu$, panel (c) shows the Pearson coefficient between these two quantities, at various $T$: both Hebb's and SK show some mutual dependence, but $\omega_i^{\mu}$ and $\Delta_i^\mu$ in the Hebbian landscape appear to be more correlated. 
Finally, panel (d) shows $\omega_{emp}(0)$ in both cases. This measure is consistent with the indication coming from the Pearson coefficient: while for the Hebb's initialization $\omega_{emp}(0)$ remains significantly negative and reaches the lowest values at low temperatures\footnote{Specifically, the lowest value of $\omega_{emp}(0)$ in the Hebbian case is reached at low, but non-zero temperature. This suggests that saddles in the energy landscape also can be good training configurations: the study of this particular case is reported in Appendix \ref{app:saddles}. }, the random case shows the opposite trend, with the estimated $\omega_{emp}(0)$ staying generally close to zero.\\ 

Going back to the Hebbian energy landscape, one can also study how the distribution of $(\Delta_i^\mu, \omega_i^{\mu})$ evolves during the training process.  Fig.~\ref{fig:omega_dynamic}a shows the value of $\omega_{emp}(0)$ at different time steps of the training process, for different values of $\alpha$, when training configurations are sampled at $T = 0$. We find that $\omega_{emp}(0)<0$ throughout training for $\alpha \leq 0.6$. The progressive increase of $\omega_{emp}(0)$ means that fixed points of the initial Hebbian landscape are better training configurations compared to fixed points at intermediate stages of training. In the last part of training, points reacquire more negative values, but this is not a reliable indication of good performance: as shown in \cref{fig:omega_dynamic}c, in this part of the process the standard deviation of the couplings $\sigma_i$ is comparable to $O(\lambda\cdot N^{-1/2})$, and the expansion of the $\mathcal{L}$ in \cref{eq:deltaL_lwsn} is not valid. The last part of the training, where $\sigma_i \simeq 0$ $\forall i$, has been neglected from the plot. 
This picture is confirmed by the study of the Pearson correlation coefficient between $\omega_i^{\mu}$ and the associated stabilities $\Delta_i^{\mu}$ (see \cref{fig:omega_dynamic},b). High values of the Pearson coefficient show a strong dependence of the structure of noise on the relative stabilities. For all $\alpha$, the Pearson coefficient is highest at $d = 0$, and progressively decreases during training, suggesting that the quality of the training configurations is deteriorating. The final increase in the coefficient is, again, due to the vanishing of the standard deviations $\sigma_i$ of the couplings, and does not indicate good performance.\\

\section{Hebbian Unlearning is Training with Structured Noise \label{sec:HU_TWN}}
\begin{figure}[t]
\centering
\includegraphics[width=0.85\linewidth]{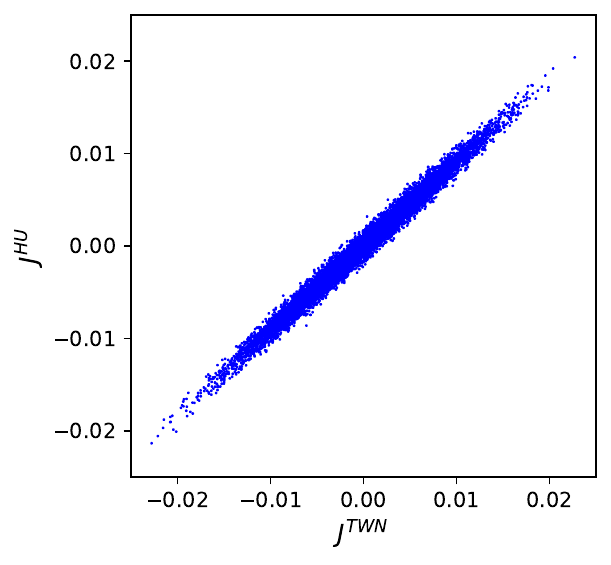}
\caption{Elements of the $J$ matrix obtained from HU $vs.$ the one resulting from TWN for one realization of the memories after the same amount of training steps. Choice of the parameters: $N = 500$, $\alpha = 0.5$, $\lambda = 5\cdot10^{-3}$. }
\label{fig:scatter}
\end{figure}

Motivated by the results of the previous Section on the characterization of effective training configurations, we implement the TWN algorithm by only using stable fixed points of the dynamics having an overlap with the memories $m_t = 0^+$. As showed by the numerics (see \cref{fig:omegatilde_SK} and \cref{fig:omega_dynamic}) such states effectively satisfy, for a large amount of algorithm steps, the optimal noise condition (\ref{eq:deltaL_lwsn3}). We prove that TWN converges to the HU rule in this particular scenario and that it is suitable to be adapted to an unsupervised learning rule.\\ \\
The total update to the couplings at time $D$ by the symmetric TWN algorithm in \cref{eq:lwn} can be decomposed as a sum of two contributions 
\begin{equation}
\label{eq:DeltaJ_TWN}
\Delta J_{ij}(D) = \Delta J_{ij}^N(D) + \Delta J_{ij}^U(D).
\end{equation}
The first, which will be referred to as \textit{noise} contribution, is expressed in terms of noise units as
\begin{equation}
\small
    \label{eq:Npart}
    \Delta J_{ij}^N(D) = \frac{\lambda}{2N}\sum_{d = 1}^D \xi_i^{\mu_d}\xi_j^{\mu_d}\chi_j^{\mu_d} + \frac{\lambda}{2N}\sum_{d = 1}^D\xi_j^{\mu_d}\xi_i^{\mu_d}\chi_i^{\mu_d},
\end{equation}
while the second, which will be referred to as \textit{unlearning} contribution, is given by
\begin{equation}
    \label{eq:Upart}
    \Delta J_{ij}^U(D) = -\frac{\lambda}{2N}\sum_{d = 1}^D \left(S_i^{1,\mu_d}S_j^{\mu_d} + S_i^{\mu_d}S_j^{1,\mu_d}\right).
\end{equation} 
Notice that, in the maximal noise case $m_t=0^+$, $\Delta J_{ij}^N(D)$ averages to zero over the process. When the number of steps $D$ is proportional to $N/\lambda$, its variance is of order $O(\lambda/N)$, leading to
\begin{equation}
\label{eq:negligible}
    \Delta J_{ij}^N(D) = O\left(\sqrt{\frac{\lambda}{N}}\right).
 \end{equation} 
 Hence, when $m_t = 0^+$ and $\lambda/N \rightarrow 0$, the noise contribution is negligible, and the TWN update rule reduces to 
\begin{equation}
\label{eq:TWN_shortened}
    \delta J_{ij}^{(d)} =-\frac{\lambda}{2N}\big(S_i^{1,\mu_d}S_j^{\mu_d} + S_i^{\mu_d}S_j^{1,\mu_d}\big).
\end{equation} 
In this limit, if fixed points of the dynamics are used for training, i.e. $\vec{S}^{\mu_d} = \vec{S}^{1,\mu_d}$, TWN and the HU converge to the same update rule.\\
Hence, the two algorithms must perform equally when we employ stable fixed points of the dynamics as training data-points. In order to test numerically this statement we proceed in the following manner: the network is initialized according to Hebb's learning rule; we train $J$ by means of: the TWN update rule (\ref{eq:lwsn}) using stable fixed points of the dynamics with overlap
\footnote{Fixed points are sampled by initializing the network with an overlap in $(0, N^{-1/2})$ with one memory, and then running the dynamics to a fixed point. Typically, such an initial overlap implies the same order of magnitude for the final overlap.} 
$m=0^+$; the HU update rule (\ref{eq:unl_rule}) using stable fixed points of the dynamics randomly sampled in the landscape (as described in section \ref{sec:unl}).
The resulting networks are compared after a fixed amount of iterations $D$, verifying that they overlap significantly. More precisely, fig. \ref{fig:scatter} compares the elements of the two matrices for one of our experiments, reporting a strong correlation between the two.

Numerically, it is known that HU performs as well as SVMs when $\alpha < \alpha_c^{HU}$ (see \cite{benedetti_supervised_2022}). Our analysis of the optimal structure of noise can explain this excellent performance. 
In fact, as showed by \cref{fig:omegatilde_SK}d, local minima of an energy landscape initialized according to Hebb's learning rule yield very negative $\omega_i^{\mu}$ when $\Delta_i^{\mu}\sim 0$, driving the network towards the SVM solution. \\
Interestingly, the weights $\omega_i^{\mu}$ have a precise geometric interpretation in the case of HU, where $\omega_i^{\mu} = m_{\mu}\chi_i^{\mu}.$ Define the following $N\text{-dimensional}$ vectors: $\vec{J}_i$ as the collection of the elements contained in the $i^{th}$ row of the connectivity matrix, $\vec{\eta}_i^\mu=\xi_i^\mu\vec{\xi}^\mu$ and $\vec{\eta_i}^{\mu_d} = S_i^{\mu_d}\vec{S}^{\mu_d}$, where $\vec{S}^{\mu_d}$ is the training configuration. 
Then, at iteration $d$, one has
\begin{equation}
\label{eq:omega_interp}
\omega_i^{\mu} = (\vec{\eta_i}^{\mu_d} \cdot \vec{\eta_i}^{\mu})/N.
\end{equation}
On the other hand, stabilities can be written as
\begin{equation}
    \Delta_i^\mu=\frac{\vec{J_i} \cdot \vec{\eta}_i^\mu }{\sqrt{N}\sigma_i},
\end{equation}
and the HU rule becomes 
\begin{equation}
    \delta\vec{J_i}^{(d)} = - \frac{\lambda}{N}\vec{\eta_i}^{\mu_d}.
\end{equation}
It follows that $\delta \Delta_i^{\mu} \propto -\omega_i^{\mu}$: a negative value of $\omega_i^{\mu}$ will increase the degree of alignment between $\vec{J_i} \text{ and } \vec{\eta}_i^\mu$, improving the stability. 
Such an anisotropic distribution of the sampled $\vec{\eta_i}^{\mu_d}$ around $\vec{\eta_i}^{\mu}$ is exactly what condition (\ref{eq:deltaL_lwsn3}) prescribes for good training configurations. Notice that this anisotropy is hidden in the space of configurations, where minima $\vec{S}^{\mu_d}$ of the energy landscape   surround memories $\vec{\xi}^{\mu}$ isotropically. The fact that $\omega_i^{\mu}$ are not i.i.d. variables (as one would have for $\chi_i^{\mu}$ generated as a consequence of \cref{eq:chi}), but rather their statistics depends $(i,\mu)$, is the real physical interpretation of the data-structure. Interestingly, considering only the transformed memories $\vec{\eta_i}^{\mu}$ that have $\Delta_i^{\mu} \sim 0$ is sufficient to drag all stabilities above zero.\\

Finally, it is essential to note that, since the training overlap under consideration is close to $0$, stable fixed points can be sampled in a full unsupervised fashion: when the dynamics is initialized at random (which implies having an overlap $m_t = 0^+$ with some memory) it will typically conserve a small overlap (with the same memory) even at convergence \cite{kepler_domains_1988}. As a consequence, the argument presented above holds, and a supervised algorithm as TWN can be reduced to a more biologically plausible and faster unsupervised learning rule, as the HU one.

\section{\label{sec:conc}Conclusions}
Gardner's training-with-noise (TWN) algorithm  \cite{gardner_training_1989} showed that noise can be injected during training to increase generalization, i.e. the capability of neural networks to retrieve memories by receiving corrupted stimuli \cite{drop, tadros, saad_learning_1996,schottky_phase_1997}. In Section~\ref{sec:twn} we showed that TWN is well described by Wong and Sherrington's calculations in \cite{wong_optimally_1990, wong_neural_1993}. This implies convergence to a Hebbian matrix or a Support Vector Machine (SVM) when learning random configurations with, respectively, maximal (i.e. $m_t = 0^+$) or minimal (i.e. $m_t = 1^-$) amount of noise. The training overlap $m_t$ can be then tuned to interpolate between these two models. \\
In Section~\ref{sec:choice_of_training_conf} we consider the maximal noise scenario represented by $m_t = 0^+$, and derived the condition to be satisfied by the noise in order to approach the performance of a SVM. Specifically, the entries of training configurations (i.e. the data features) are constrained by internal dependencies that we call \textit{structure}. We have explored the landscape of attractors numerically, and showed that in the lower levels of a Hebbian energy landscape there exist a band of maximally noisy saddle states which are excellent training data. Specifically such states, among which we also find local minima of the energy, are able to train a network which shares the same properties of a SVM. \\ 
In Section~\ref{sec:HU_TWN} we proved that, when stable fixed points of the dynamics are used as training data, the TWN algorithm and the HU algorithm are equivalent. This sheds light over why basins of attraction generated by HU are extremely large, as shown in \cite{benedetti_supervised_2022}: training data are such as to approach the global minimum of the loss function of a SVM, which is known in literature to have high associativity properties. \\

As neuroscientists become more convinced of the importance of sleep in memory consolidation \cite{creery_electrophysiological_2022, maingret_hippocampo-cortical_2016}, connections arise between unsupervised training algorithms on machines and synaptic plasticity processes that occur outside the wakeful hours \cite{crick_function_1983, girardeau_brain_2021, hinton_forward-forward_2022, hinton_unsupervised_1999}. In light of these remarks and the work presented in this article, one might conceive natural learning as a \textit{two-phase} process. During a first \textit{online} phase, external stimuli are processed by the network through the standard TWN algorithm. One can imagine the stimuli to be maximally noisy versions of some unknown \textit{archetypes} embedded into the environment. As a consequence, training will shape a pure Hebbian landscape of attraction out of the retrieval regime. In a second \textit{offline} phase, the early formed network samples structured noisy neural configurations, still weakly correlated with the archetypes, from the landscape of attractors. Such states could be, for instance, stable fixed points of the neural dynamics. These neural configurations are then processed by the same kind of TWN algorithm, and memory is consolidated by centering the unconscious archetypes in the middle of large basins of attraction.\\ 
In this context, the results contained in this work make progress on two fronts. On one side, they shed light over the structure of noise that is optimal for learning in neural networks, helping to develop a finer theory behind the very empirical techniques of noise injection implemented in training deep networks \cite{drop, shorten_survey_2019, tomasini_how_2023, bonnasse-gahot_categorical_2022}. On the other side, they draw a connection between unsupervised, and thus more biologically relevant, learning processes and the supervised ones, from which most of the modern theory of neural networks derive. 

\begin{acknowledgments}
The authors are particularly grateful to their mentors Enzo Marinari, Giancarlo Ruocco and Francesco Zamponi for precious suggestions and support. They also thank Fabian Aguirre Lopez, Aldo Battista, Simona Cocco, Giampaolo Folena, Rémi Monasson and Mauro Pastore for useful discussions, as well as an anonymous referee for helpful comments that have significantly improved the manuscript. 
\end{acknowledgments}

\bibliographystyle{unsrt}
\bibliography{biblio}
\newpage

\appendix
\onecolumngrid

\section{Descending over the $\mathcal{L}$ function}
\subsection{Training with noise}
\label{sec:appA1}
At each step of the algorithm, a memory label $\mu_d$ is sampled at random and the update (\ref{eq:lwn}) is performed over the couplings. The new value of the $\mathcal{L}$ function (\ref{eq:loss_ws}) is
\begin{equation}
    \label{eq:new_loss}
    \mathcal{L}^{'} = -\frac{1}{\alpha N^2}\sum_{i,\mu }^{N,p}\text{erf}\left(\frac{m\Delta_i^{\mu}}{\sqrt{2(1-m^2)}} + \frac{\lambda m}{N\sigma_i\sqrt{2N(1-m^2)}}\epsilon_i^{\mu_d}\xi_i^{\mu}\xi_i^{\mu_d}\sum_{j \neq i}S_j^{\mu_d}\xi_j^{\mu}\right),
\end{equation}
where 
\begin{equation}
    \epsilon_i^{\mu_d} = \Theta\left(\xi_i^{\mu_d}\sum_{k=1}^N J_{ik}S_k^{\mu_d} \right). 
\end{equation}
Since $\delta\sigma_i \propto \frac{\lambda}{N}\frac{J_i}{\sigma_i} + O\left(\frac{\lambda}{N} \right)^3$ and the mean $J_i$ of the couplings along line $i$ equals zero by initialization and it is naturally maintained null during the algorithm, we have considered 
$\sigma_i^\prime\simeq \sigma_i$. 
Then $\mathcal{L}'$ can be rewritten as
\begin{equation}
\label{eq:new_loss2}
    \begin{aligned}  
    \mathcal{L}^{'} = &-\frac{1}{\alpha N^2}\sum_{i,\mu\neq \mu_d}^{N, p}\text{erf}\left(\frac{m\Delta_i^{\mu}}{\sqrt{2(1-m^2)}} + O\left( \frac{1}{N}\right)\right) -\\
    &-\frac{1}{\alpha N^2}\sum_{i}^{N}\text{erf}\left(\frac{m\Delta_i^{\mu_d}}{\sqrt{2(1-m^2)}} + \frac{\lambda m\cdot m_t}{\sigma_i\sqrt{2N(1-m^2)}}\epsilon_i^{\mu_d} + O\left(\frac{1}{N}\right)\right),
    \end{aligned}
\end{equation}
where we have used that $\frac{1}{N}\sum_{j\neq i}\xi_j^{\mu}S_j^{\mu_d} = O(N^{-1/2})$ when $\mu \neq \mu_d$ and $m_t = O(1)$ . We thus expand the errorfunction at the first order in $O(N^{-1/2})$ obtaining the variations to $\mathcal{L}$ in equation (\ref{eq:delta_loss}). \\
As a technical comment, note that standard deviation along one row of the couplings matrix $\sigma_i$ (see \cref{eq:stab}) is a variable quantity over time, and numerics suggest it is slowly decreasing. As a result, the expansion performed to determine the variation of $\mathcal{L}$ (see \cref{eq:new_loss2} in Appendix \ref{sec:appA1}) might not be justified after a certain number of steps, leading to a non-monotonic trend of the loss function. The non-monotonic trend of $\mathcal{L}(m_t, J(m_t))$ due to this effect is shown in the inset of \cref{fig:loss}. However, this inconvenience can be overcome by rescaling the learning rate $\lambda$ into $\lambda_i = \lambda\cdot \sigma_i$ at each iteration, obtaining the curves in \cref{fig:loss}.
\begin{figure}[ht!]
\centering
\includegraphics[width=0.55\linewidth]{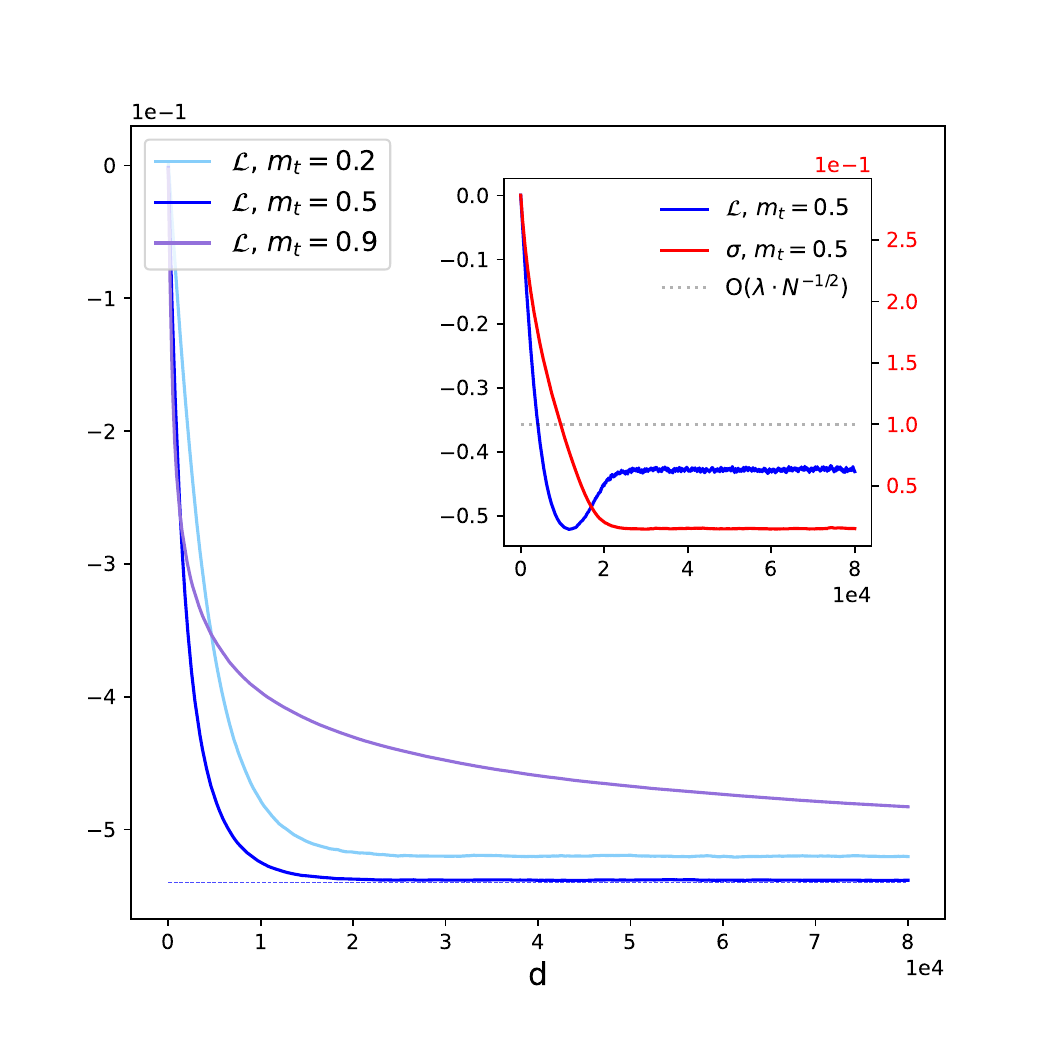}
\caption{ The \textit{blueish} lines in the main plot report the function $\mathcal{L}(m = 0.5, J(m_t))$ for different training overlaps as functions of the number of algorithm steps $d$. The \textit{dotted} line represents the theoretical minimum value from \cite{wong_neural_1993}. The learning strength $\lambda$ has been rescaled by the standard deviation of the couplings as described in the text. The subplot reports the case $m_t = 0.5$ when the learning strength is not rescaled: $\mathcal{L}$ is in \textit{blue}, while a measure of the standard deviation of the couplings, defined as $\sigma = \frac{1}{N}\sum\sigma_i$, is reported in \textit{red}:  the value $\lambda\cdot N^{-1/2}$ of the standard deviation is depicted in \textit{light gray} to properly signal the moment when equation (\ref{eq:delta_loss}) loses its validity. All measures are averaged over $5$ realizations of the couplings $J$. Choice of the parameters: $N = 100$, $\alpha = 0.3$, $\lambda = 1$, the initial couplings are Gaussian with unitary mean, zero variance and $J_{ii}^{(0)} = 0$ $\forall i$.}
\label{fig:loss}
\end{figure}
\subsection{Training with structured noise}
\label{sec:appA2}
\begin{figure*}[ht!]
\begin{tabularx}{\linewidth}{*{2}{X}}
     \begin{subfigure}[b]{\linewidth}
         \centering
         \includegraphics[width=0.9\textwidth]{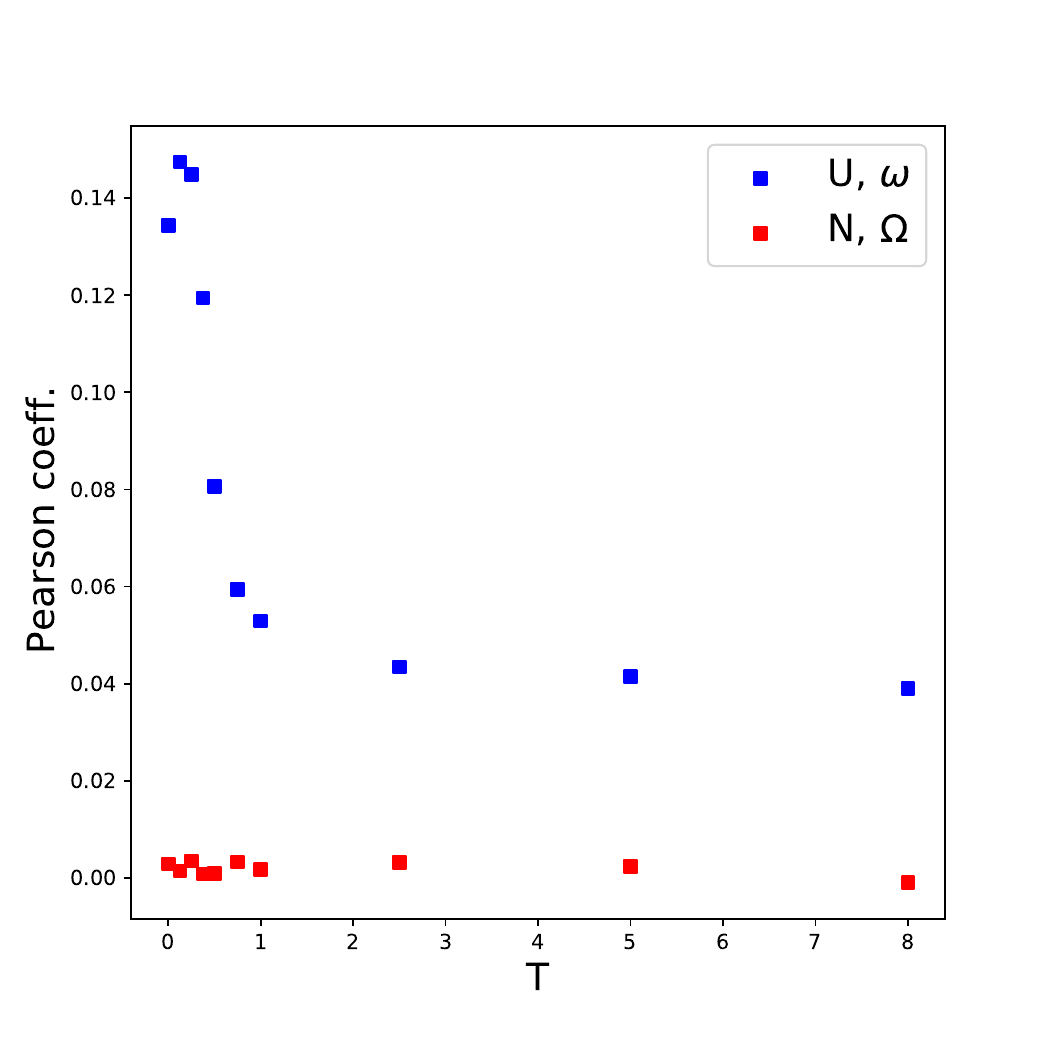}
         \caption{}
         \label{fig:ptest_a}
     \end{subfigure}
     &
     \hfill
     \begin{subfigure}[b]{\linewidth}
         \centering
         \includegraphics[width=0.9\textwidth]{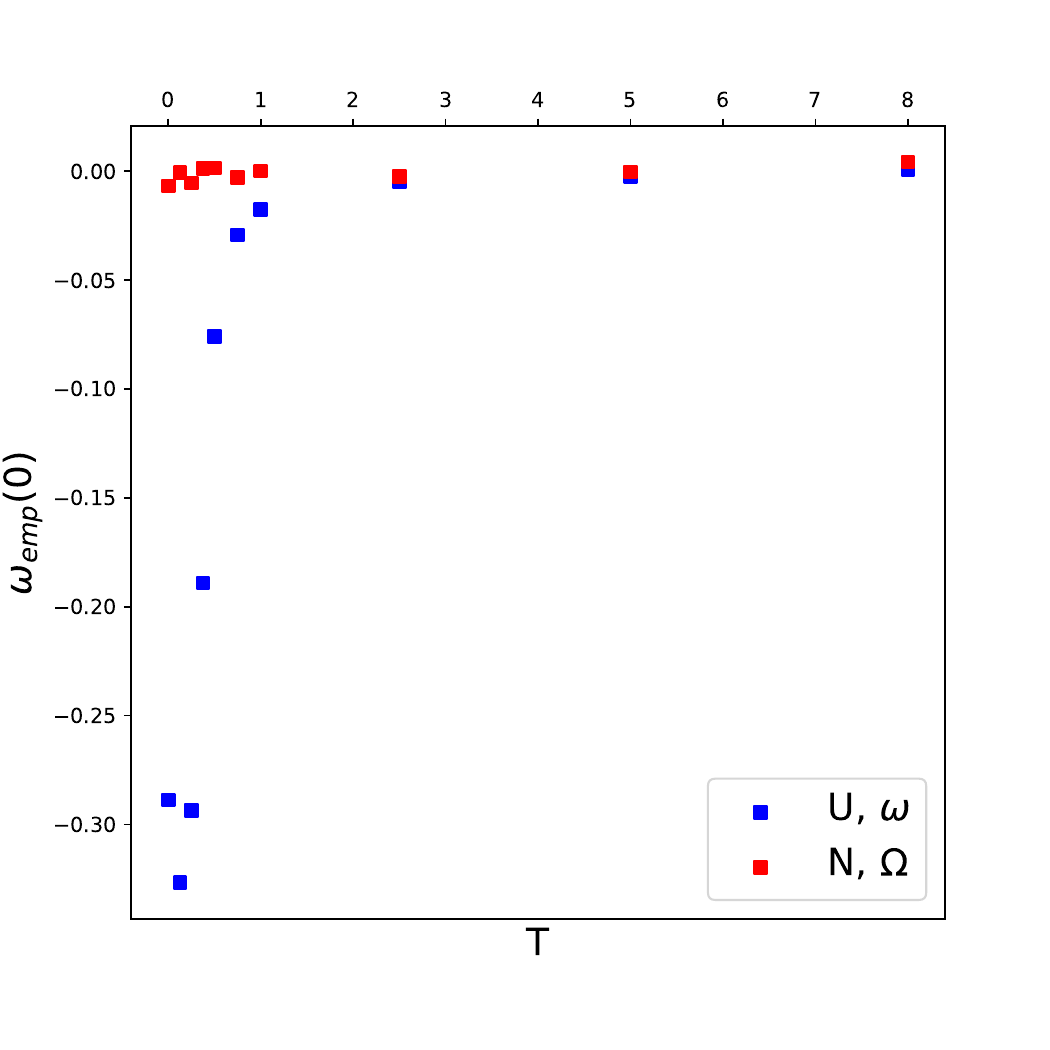}
         \caption{}
         \label{fig:ptest_b}
     \end{subfigure}
\end{tabularx}
\caption{(a) Pearson coefficient between $\omega_i^{\mu}$, $\Omega_i^{\mu}$ and the stabilities $\Delta_i^{\mu}$, (b) $\omega_{emp}(0)$ for the case of a Hebbian network at different temperatures $T$. Configurations at a given temperature $T$ have been sampled by a Monte Carlo of the Kawasaki kind, in order to choose only maximally noisy states ($m_t = 0^+$). Points are collected from $15$ samples of the network. Choice of the parameters: $N = 500$, $\alpha = 0.5$.}
\end{figure*}

The new value of $\mathcal{L}$ is derived by using equation (\ref{eq:lwn2}) to evaluate the variation of stabilities
\begin{equation}
    \begin{aligned}
    \label{eq:new_lossApp}
    \mathcal{L}^{'} = -\frac{1}{\alpha N^2}\sum_{i,\mu }^{N,p}\text{erf}\Bigg(&\frac{m\Delta_i^{\mu}}{\sqrt{2(1-m^2)}} + \frac{\lambda m \xi_i^{\mu}\xi_i^{\mu_d}}{2N\sigma_i\sqrt{2N(1-m^2)}}\sum_{j=1}^N S_j^{\mu_d}\xi_j^{\mu} + \frac{\lambda m \xi_i^{\mu} S_i^{\mu_d}}{2N\sigma_i\sqrt{2N(1-m^2)}}\sum_{j=1}^N \xi_j^{\mu_d}\xi_j^{\mu} -\\ 
    -&\frac{\lambda m \xi_i^{\mu}S_i^{1,\mu_d}}{2N\sigma_i\sqrt{2N(1-m^2)}}\sum_{j=1}^N S_j^{\mu_d}\xi_j^{\mu} - \frac{\lambda m \xi_i^{\mu}S_i^{\mu_d}}{2N\sigma_i\sqrt{2N(1-m^2)}}\sum_{j=1}^N S_j^{1,\mu_d}\xi_j^{\mu}\Bigg),
    \end{aligned}
\end{equation}
where $\sigma_i^{'} \simeq \sigma_i$ as in the previous paragraph. We now redefine $\chi_i^{\mu} = \xi_i^{\mu}S_i^{\mu_d}$,  $\chi_i^{1,\mu} = \xi_i^{\mu}S_i^{1,\mu_d}$, $m_{\mu} = \frac{1}{N}\sum_{j = 1}^N S_j^{\mu_d}\xi_j^{\mu}$ and $m_{1,\mu} = \frac{1}{N}\sum_{j = 1}^N S_j^{1,\mu_d}\xi_j^{\mu}$ and expand the errorfunction at the first order in $O(N^{-1/2})$ obtaining
\begin{equation}
\label{eq:deltaL_tot}
\delta\mathcal{L} = \frac{m\lambda}{\sqrt{2\pi\alpha^2 N^5 (1-m^2)}}\sum_{i,\mu}^{N,p}\frac{1}{2\sigma_i}\left[(m_{\mu}\chi_i^{1,\mu} + m_{1,\mu}\chi_i^{\mu}) - (m_{\mu}\xi_i^{\mu}\xi_i^{\mu_d} + M_{\mu}^{\mu_d}\chi_i^{\mu})\right] \exp{\left(-\frac{m^2\Delta_i^{\mu^2}}{2(1-m^2)}\right)},
\end{equation}
where $M_{\mu}^{\mu_d} = \frac{1}{N}\sum_{i=1}^N \xi_i^{\mu}\xi_i^{\mu_d}$. Equation (\ref{eq:deltaL_tot}) can be decomposed in
\begin{equation}
    \delta \mathcal{L} = \delta \mathcal{L}_{N} + \delta \mathcal{L}_{U},
\end{equation}
where $\delta \mathcal{L}_{U}$ contains the weight 
\begin{equation}
    \label{eq:w1}
    \omega_i^{\mu} =\frac{1}{2\sigma_i}\left( m_{\mu}\chi_i^{1,\mu} + m_{1,\mu}\chi_i^{\mu}\right),
\end{equation}
while $\delta \mathcal{L}_{N}$ contains 
\begin{equation}
    \label{eq:w2}
    \Omega_i^{\mu} =\frac{1}{2\sigma_i}\left(m_{\mu}\xi_i^{\mu}\xi_i^{\mu_d} + M_{\mu}^{\mu_d}\chi_i^{\mu}\right).
\end{equation}
We study the two contributions numerically, on a Hebbian network, i.e. with no learning going on, for the case of $m_t = 0^+$. The Pearson coefficient is measured between the vector of the stabilities $\Delta_i^{\mu}$ and the weights $\omega_i^{\mu}$ as well as with  $\Omega_i^{\mu}$ separately. This quantity should underline an eventual reciprocal dependence between $\omega_i^{\mu}, \Omega_i^{\mu}$ and $\Delta_i^{\mu}$. The test is repeated over states sampled by a Monte Carlo at different temperatures $T$.
Results are reported in \cref{fig:ptest_a}
where it is evident that $\Omega_i^\mu$ does not have any correlation with $\Delta_i^\mu$, while the dependence of $\omega_i^\mu$ on the stabilities is evident. Moreover, we measured the indicator $\omega_{emp}(0)$, signaling the typical values of the weights $\omega_i^{\mu}$ and $\Omega_i^{\mu}$ when $\Delta_i^{\mu} \sim 0$ (see Section~\ref{sec:choice_of_training_conf} for further details), as reported in \cref{fig:ptest_b}. The plot clearly shows that $\Omega_i^{\mu}$ is small and generally fluctuating around zero. These aspects hold during the training procedure also, as it can be observed by performing the same measure at different step of the TWN procedure over states with $m_t = 0^+$. 
We will thus refer to $\delta \mathcal{L}_U$ as the relevant contribution to the variation of the function $\mathcal{L}$.

\section{\label{app:b}Training with noise - Classification analysis}
We now resume the computations performed in \cite{wong_optimally_1990, wong_neural_1993} to analyze the classification performance of TWN. We restrict ourselves to a spherical space of interactions such that 
\begin{equation}
    \sum_{j\neq i}J_{ij}^2 = N\hspace{0.7cm}\forall i,
\end{equation} 
We want to compute the probability distribution of the stabilities that now become 
\begin{equation}
\label{eq:stabApp}
    \Delta_i^{\mu} = \xi_i^{\mu}\sum_{j = 1}^N \frac{J_{ij}}{\sqrt{N}}\xi_j^{\mu}.
\end{equation}
The partition function of the model is then given by
\begin{equation}
\label{eq:Z}
    Z = \int\prod_{i,j}dJ_{ij}\delta\left(\sum_{j\neq i}J_{ij}^2 - N\right)\exp{\Big(-\beta\mathcal{L}(m_t,J)\Big)},
\end{equation}
where $\beta$ is the inverse annealing temperature of the problem and the loss function $\mathcal{L}$ is defined in \cref{eq:loss_ws}. The distribution of the stabilities is
\begin{equation}
    \small
    \label{eq:rho1}
    \rho_{m_t}(\Delta) = \overline{\frac{1}{Z}\int \prod_{j}dJ_{\bullet j}\delta\left(\sum_{j}J_{\bullet j}^2 - N\right)\exp{\left(\beta\sum_{\mu}\text{erf}\left(\frac{m_t\xi_\bullet^{\mu}\sum_jJ_{\bullet j}\xi_j^{\mu}}{\sqrt{2N(1-m_t^2)}}\right)\right)}\delta\left( \xi_\bullet^1\sum_j\frac{J_{\bullet j}}{\sqrt{N}}\xi_j^1-\Delta\right)},
\end{equation}
where $\overline{\hspace{0.1cm}\cdot\hspace{0.1cm}}$ denotes the average over the realizations of the memories and we have neglected the factorization over $i$, since we treat the optimization process as independent along the lines of the $J_{ij}$ matrix. Index $i$ has been nevertheless substituted by $\bullet$ for the sake of completeness.  
Replicas can be used to evaluate the normalization, i.e.
\begin{equation}
    1/Z = \lim_{n\rightarrow 0}Z^{n-1}.
\end{equation}
The replica calculation in the replica symmetric ansatz \cite{mezard_spin_1986} when $\beta \rightarrow \infty$ leads to the following expression for the density of the stabilities
\begin{equation}
    \label{eq:rho}
    \small
    \rho_{m_t}(\Delta) = \frac{1}{\sqrt{2\pi}}\left(1  + \sqrt{\frac{2}{\pi}}\frac{m_t^3 \chi}{(1 - m_t^2)^{3/2}}\Delta\exp\left( -\frac{m_t^2\Delta}{2(1-m_t^2)}\right)\right)\exp{-\frac{w(\Delta)^2}{2}},
\end{equation}
where $\chi$ and $w$ are derived by solving the following two equations
\begin{equation}
\label{eq:sad1}
    w = x - \frac{\sqrt{2}m_t\chi}{\sqrt{\pi(1 - m_t^2)}}\exp{\left(-\frac{m_t^2x^2}{2(1-m_t^2)}\right)},
\end{equation}
\begin{equation}
\label{eq:sad2}
    \int_{-\infty}^{+\infty} Dw\left(x^*(w,\chi) - w\right)^2 = \alpha^{-1},
\end{equation}
with $x^*$ being the solution of (\ref{eq:sad1}) and $Dw$ being the standard Gaussian measure with zero mean and unitary variance.

\section{\label{app:c}Training with noise - Generalization analysis}
\begin{figure}[ht!]
\centering
\includegraphics[width=0.5\linewidth]{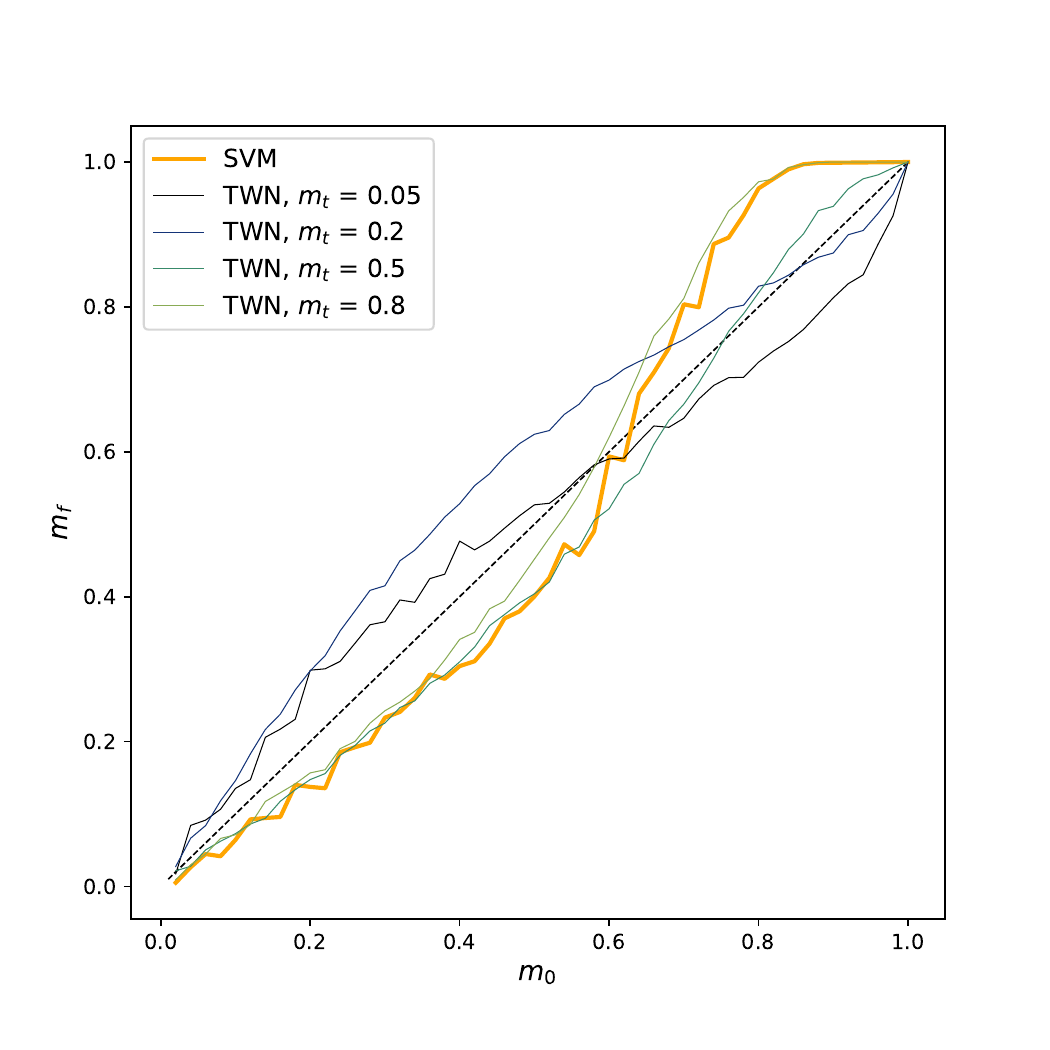}
\caption{Retrieval map $m_f(m_0)$ in the case of networks trained through SVM and TWN algorithms. Curves are shown as a function of $m_t$ and compared with the bisector, indicated with a \textit{dashed black} line. Points are averaged over $10$ samples. Choice of the parameters: $N = 200$, $\alpha = 0.45$.}
\label{fig:mfmi_evo}
\end{figure}

We report here a general experimental procedure to measure the average size of the basins of attraction of a fully connected neural network of finite size $N$ and a given choice of the control parameters.\\ The network is firstly trained according to an algorithm of our choice. Once the couplings have been found, the asynchronous version of dynamics (\ref{eq:dynamics}) is initialized in one of the memories. The dynamics is run until convergence onto the attractor associated to the basin of belonging of the memory. Now the retrieval map $m_f(m_0)$ is measured with respect to that particular attractor, and the procedure is repeated over different memories and realizations of the network. The average radius of the basin of attraction is then measured as the value of $1-m_0$ where $m_f(m_0)$ equals a reference value. In our case such value is $m_f = 0.98$. 
We have applied this procedure on networks trained either as SVMs and with the TWN algorithm. In the former case a convex algorithm contained in the $cvxpy$ Python domain \cite{diamond_cvxpy_nodate} is implemented to train the network. To be more specific, $N$ independent machines are trained to correctly classify $p = \alpha N$ binary memories of the kind of $\vec{\xi}^{\mu} \in \{-1,+1\}^N$ having as labels $\xi_i^{\mu}$ with $i \in [1,..,N]$.
\\Regarding the dynamics, the stability of fixed points is in general implied by some properties of the couplings, mainly their degree of symmetry. For the case of the TWN algorithm we start from a random symmetric matrix, as done in \cite{kepler_domains_1988}: the update of the couplings will only perturb the initial symmetry yet allowing the measures to be still consistent with the theory. 
On the other hand, numerics show that SVMs are sufficiently symmetric to let the asynchronous dynamics converge. The comparison between the retrieval maps obtained for the two algorithms with $\alpha = 0.45$ and $N = 200$ is reported in \cref{fig:mfmi_evo}. The curve relative to the SVM is fixed, while the one associated with the TWN is changing with respect to the training overlap $m_t$.
\\Even if the SVM always reaches classification of the memories for $\alpha < 2$ \cite{gardner_space_1988}, a network trained with noise might show two different behaviors: one associated to \textit{retrieval} where each memory is close to an attractor, and one related to \textit{non-retrieval} where the memory is far from its attractor and the basins of attraction might contain orthogonal configurations with respect to the central attractor. 
In particular network models where couplings are assembled according to particular rules (e.g. \cite{hopfield_neural_1982,dotsenko_statistical_1991}) the transition between these two regimes can be computed analytically. 
In the case of TWN this cannot be done. It is then important to find an empirical criterion to divide the two behaviors as a function of $(m_t,\alpha)$.
Let us assume that when $N \gg 1$ the retrieval map $m_f(m_0)$ develops a plateau starting from $m_0 = 1$ and ending in some limit value $m_0 = m_c < 1$ such that $m_f = 1$ along all this interval (as in \cref{fig:mfmi_evo}). Hence, one can associate the formation of such a plateau with the existence of a cohesive basin of attraction, where close configurations in hamming distance to the attractor converge to the attractor. One then wants to measure the value of $m_t$ at which such property of the basin disappears. As a possible estimate, it is convenient to consider the $m_t^*$ such that
\begin{equation}
\label{eq:change_conv}
    \frac{dm_f}{dm_0}(m_t^*)\Big|_{m_0 = 1} =  1.
\end{equation}
The numerical extrapolation of the overlap in \cref{fig:mf1} at different values of $N$ shows a good agreement between the approximate separation between the two regions showed in \cref{fig:mf1} and the line estimated by condition (\ref{eq:change_conv}).

\section{Training with Saddles \label{app:saddles}}
\begin{figure}[ht!]
\centering
\includegraphics[width=0.7\linewidth]{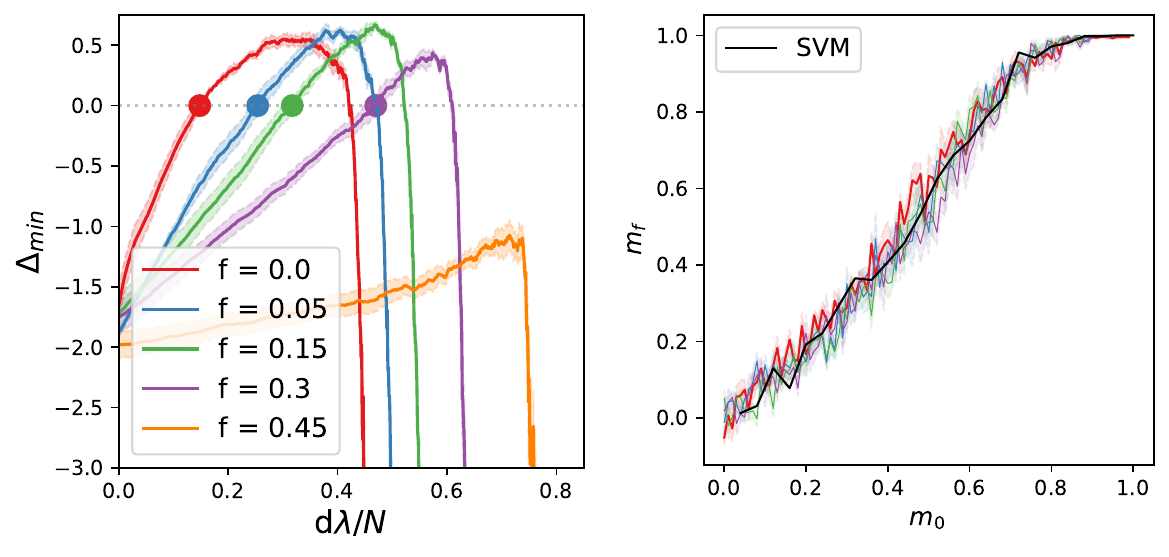}
\caption{Left: Minimum stability $\Delta_{\text{min}} = \min_{i,\mu}(\Delta_i^{\mu})$ as a function of the algorithm steps on a network trained with the TWN routine that learns saddles of various indices $f$. The initial matrix is assembled according to the Hebb's rule. Full dots report the amount of iterations needed to accomplish classification. Right: the retrieval map $m_f(m_0)$ as measured on the positions of the colored dots from the right panel, with the same color code being applied. A comparison with a SVM trained with the same choice of the parameters is also presented through the \textit{dashed blue} line. All measures are averaged over $5$ samples with the shaded region indicating the experimental errors. The choice of the parameters is: $N = 100$, $\alpha = 0.35$, $\lambda = 10^{-3}$.}
\label{fig:stab_sads}
\end{figure}

Notice, from both panels (c), (d) of \cref{fig:omegatilde_SK}, the existence of an optimum which does not coincide with the stable fixed points of the dynamics (i.e. $T = 0$). As noted in previous studies on spin glasses \cite{aspelmeier_free_2006}, one can associate configurations probed by a Monte Carlo at finite temperatures with configurations which are typically saddles in the energy landscape with a given $\textit{saddle index}$. The $\textit{saddle index}$ is defined as the ratio between the number of unstable sites under the dynamics (see \cref{eq:dynamics}) and the total number of directions $N$. Stable fixed points have $f = 0$, while random configurations are expected to have $f = 1/2$. In order to check whether a particular $f$ is capturing relevant features of the virtuous training configurations, we sampled training data according to the requirement that their saddle fraction assumes a specific value $f$ and $m_t=0^+$. Saddles are then employed for training the network according to \cref{eq:lwn}. Sampling is performed by randomly initializing the network on a configuration having training overlap $m_t = 0^+$ with a reference memory, and performing a zero temperature dynamics on the landscape defined by the new energy
\begin{equation}
    \label{eq:sad_energy}
    \mathcal{E}(\vec{S}|f, J) = \frac{1}{2}\left(\frac{1}{N}\sum_{i=1}^N\Theta(-S_i\sum_{j = 1}^N J_{ij}S_j) - f\right)^2,
\end{equation}
where $\Theta(x)$ is the Heaviside function. Yet again, the value of $m_t$ was maintained constant during the descent. 
The left panel in \cref{fig:stab_sads} shows how the minimum stability evolves during the training process while a TWN algorithm is initialized in the Hebbian matrix and learns saddles of different indices. For a network of $N = 100$ and $\alpha = 0.35$, we found that classification (i.e. positive minimum stability) is reached until a certain value of $f$, suggesting that saddles belonging to this band are indeed good training data. The band of saddles that are suitable for learning shrinks when $\alpha$ increases until such states do not significantly satisfy \cref{eq:deltaL_lwsn3} anymore. Such limit capacity is located around the critical one for HU. 
It should be stressed that the precise performance as a function of $f$ is quite sensitive to the sampling procedure. 
Simulated annealing routines \cite{s_optimization_1983} have also been employed to minimize (\ref{eq:sad_energy}), obtaining qualitatively similar results yet not coinciding with the ones reported in \cref{fig:stab_sads}. A qualitative study of the basins of attraction of the network has been performed and reported in the right panel in \cref{fig:stab_sads}. 
Specifically, the retrieval map $m_f(m_0)$ has been measured relatively to the saddle indices $f$ at the first time they reached classification, in analogy to what has been measured in \cite{benedetti_supervised_2022}. 
The curves coincide quite well, suggesting that finite sized networks trained with different $f$ assume similar volumes of the basins of attraction when they are measured at the very first instant they reach classification. The plot also shows that the generalization performance is comparable with the one of a SVM trained with the same choice of the control parameters.\\

\end{document}